\documentclass[review]{elsarticle}
\usepackage{bm}
\usepackage{amsmath,amsthm,amssymb,graphicx,tikz,esint}
\usepackage[toc,page]{appendix}
\usepackage{epstopdf}
\usepackage{float}
\usepackage{subcaption}
\usepackage{lineno}
\usepackage{hyperref}

\newcommand{\mat}[1]{ \overline{\mathbf #1} }
\renewcommand{\v}[1]{{{\mathbf #1}}}
\renewcommand{\c}[1]{{{\mathcal #1}}}

\def\dyadg #1{{\boldsymbol{\overline{#1}}}}

\modulolinenumbers[5]

\journal{Journal of Computational Physics}

\begin{document}

\begin{frontmatter}

\title{Numerical Electromagnetic Frequency Domain Analysis with Discrete Exterior Calculus}

%% Group authors per affiliation:
% \author{Elsevier\fnref{myfootnote}}
% \address{Radarweg 29, Amsterdam}
% \fntext[myfootnote]{Since 1880.}

%% or include affiliations in footnotes:
\author[mymainaddress]{Shu Chen}
\ead{shuchen5@illinois.edu}

\author[mysecondaryaddress]{Weng Cho Chew\corref{mycorrespondingauthor}}
\cortext[mycorrespondingauthor]{Corresponding author.}
\ead{w-chew@illinois.edu}

\address[mymainaddress]{Department of Physics, University of Illinois at Urbana-Champaign, IL, USA}
\address[mysecondaryaddress]{Department of Electrical and Computer Engineering, University of Illinois at Urbana-Champaign, IL, USA}

\begin{abstract}
In this paper, we perform a numerical analysis in frequency domain for various electromagnetic problems based on discrete exterior calculus (DEC) with an arbitrary $2$-D triangular or $3$-D tetrahedral mesh. We formulate the governing equations in terms of DEC for $3$D and $2$D inhomogeneous structures, and also show that the charge continuity relation is naturally satisfied. Then we introduce effective signed dual volume to incorporate material information into Hodge star operators and take into account the case when circumcenters fall outside triangles or tetrahedrons, which may lead to negative dual volume without Delaunay triangulation. Then we demonstrate the implementation of various boundary conditions, including perfect magnetic conductor (PMC), perfect electric conductor (PEC), Dirichlet, periodic, and absorbing boundary conditions (ABC) within this method. An excellent agreement is achieved through the numerical calculation of several problems, including homogeneous waveguides, microstructured fibers, photonic crystals, scattering by a $2$-D PEC, and resonant cavities.
\end{abstract}

\begin{keyword}
Maxwell's equations, differential forms, discrete exterior calculus, arbitrary simplicial mesh, circumcenter/Voronoi dual, Hodge star
\end{keyword}

\end{frontmatter}

\linenumbers

\section{Introduction}
\label{sec1}
Finite difference time domain method (FDTD) and finite element methods (FEM) have been widely applied to solve electromagnetic problems \cite{chew1995}. And they are both based on the vectorial version of Maxwell's equations. As is well known, differential forms can be used to recast Maxwell's equations in a more succinct fashion, which completely separate metric-free and metric-dependent parts, see \cite{deschamps1981, warnick1997}. Maxwell's equations and charge continuity condition in terms of differential forms within frequency domain are written as:
\begin{align}
d E = i \omega B,   \quad d H = -i \omega D + J, \quad d B = 0, \quad d D = \rho, \quad d J = i \omega \rho. 
\label{eqn:diff}
\end{align}
In this set of equations, $E$ and $H$ are $1$-forms; $D$, $B$ and $J$ are $2$-forms; $\rho$ is the only $3$-form. Electric potential $\phi$, which is not shown above, is a $0$-form. Operator $d$ is the exterior derivative, and it takes $k$-form to $(k+1)$-form. Intuitively, $k$-form is an integrand, which can be integrated over $k$-D space. For example, a line integral of $1$-form $E$\footnote{Generally speaking, with a $k$-form $\omega$ and $k$-D oriented domain $\c D$, we use $\int_{\c D} \omega$ to denote the integral of $\omega$ over $\c D$.}, $\int_a^b E = \int_a^b \v E \cdot d\v l$, leads to the potential difference from point $a$ to $b$. Indeed, the calculus of differential forms has significant advantages in illustrating the theory of EM theory compared to traditional vector analysis, see \cite{deschamps1981, warnick1997, teixeira1999}.

% It should be pointed out that, although mathematically, $\v{E}$, $\v{D}$, $\v{H}$, $\v{B}$ can all be integrated over both $1$-D space and $2$-D space to obtain a scalar value. But not both of them have physical meanings. For example, integral of $\v{E}$ field along a certain path ($1$-D) gives the potential difference, integral of $\v{B}$ field through a surface ($2$-D) leads to the magnetic flux, and integral of $\rho$ over a volume ($3$-D) is the total charge enclosed. All of these 3 integrals above can be understood as integral of corresponding forms, $\v{E}$, $\v{B}$, and $\rho$ are treated as $1$-form, $2$-form, and $3$-form, respectively.

%% add ref on the 1st line to DEC, caltech's group
Discrete exterior calculus (DEC) provides a numerical treatment of differential forms \cite{desbrun2005,desbrun2008}, which means that Equation (\ref{eqn:diff}) can be solved directly. In fact, since we are only able to obtain discrete values in numerical calculation, instead of solving for exact forms, the integral of unknown forms on finite line, area or volume are formulated and solved in DEC. Compared to FDTD, DEC can be implemented on unstructured simplicial mesh as in FEM, e.g.\ triangular mesh in $2$-D and tetrahedral mesh in $3$-D. It should be noted that DEC can be implemented on any mesh as long as circumcenter dual exists, such as regular Yee grid and hexahedral (layered triangular) grid. Therefore, this method is more adaptable over complex structures. In fact, the FDTD method can also be viewed as DEC method on Yee grid. In FDTD, the vectorial fields in fact are average values over edges, or square faces. These average values can also be thought of as small integrals, which is the origin of finite integration technique (FIT) or finite volume technique \cite{madsen1990,weiland2001}. Moreover, in contrast to FEM, this method exactly preserves important structural features of Maxwell's equations, e.g.\ Gauss's law $\nabla \cdot \v{D} = \rho$ and $\v{E} = -\nabla \phi$. Besides, since $\nabla \times \nabla = 0$ and $\nabla \cdot \nabla \times = 0$ are naturally and exactly preserved in DEC, which means that this method will not give rise to spurious solutions due to spurious charge \cite{chew1994, na2016}. 

It should be pointed out that, the numerical work with unstructured grid based on FIT mainly dealt with $2$-D problems \cite{taflove2005, gedney1996}. More importantly, the material matrix (Hodge star) needs special treatment to be symmetric \cite{gedney2000}. The reason is that FIT uses barycentric dual; then neighboring field values are needed in interpolation for building constitutive relation. In contrast, DEC adopts circumcenter dual, or Voronoi dual, which leads to the orthogonality of primal and dual elements. Therefore, the Hodge star operators, which represent constitutive relations, are diagonal matrices (for isotropic material or anisotropic material with diagonal $\dyadg{\epsilon}$ and $\dyadg{\mu}$). Since the Hodge star operators, the metric dependent parts, are closely related to the value of $\dyadg{\epsilon}$ and $\dyadg{\mu}$, on the other hand, a novel design of $\dyadg{\epsilon}$ and $\dyadg{\mu}$ can equivalently change the metric of system, which is merit of transformation optics \cite{pendry2006}. There have been some efforts to apply DEC to computational electromagnetics. However, some limitations, mainly in two aspects, still remains in these work. First, the mesh is not totally unstructured. For example, there has been work based on polyhedral mesh \cite{he2005, rabina2014,stern2015} (layered triangular mesh for $3$-D) and partly structured nonuniform grids \cite{rabina2015}. These special meshes largely limit the use of DEC method. Second, the previous calculations are mainly in time domain with homogeneous medium. 

In our present work, we formulate and numerically calculate several kinds of electromagnetic problems in frequency domain, such as inhomogeneous waveguides (microstructured fibers), photonic crystals, and inhomogeneous resonant cavities. We also show that the charge continuity relation is exactly preserved with DEC, which prevents the generation of spurious charge and is nontrivial in FEM simulation \cite{pinto2014}. Effective signed dual volume is introduced to incorporate material information to construct Hodge star operators with arbitrary simplicial mesh. It should be pointed out that if Delaunay triangulation is performed, the positivity of dual volumes can be guaranteed \cite{hirani2013}, which leads to a positive definite Laplacian operator in free space. But for frequency domain analysis, this is not a forced requirement. We also illustrate the construction and implementation of various boundary conditions in frequency domain, including perfect magnetic conductor (PMC), perfect electric conductor (PEC), first-order and second-order absorbing boundary conditions (ABCs), and periodic boundary conditions.  

In Section 2, the framework of DEC is introduced and the governing equation for $3$-D and $2$-D problems are also formulated. Then in Section 3, we present in detail how we construct the Hodge star operators. In Section 4, various boundary conditions in frequency domain are discussed in detail. After every element of Maxwell's equations is discussed and presented with the language of DEC, some numerical examples are shown in the last section. We apply this method to solve the decoupled or coupled modes in waveguides or optical fibers, the band diagram of a photonic crystal, and resonant frequencies of $3$-D inhomogeneous resonators with both closed and open boundary. 

\section{Maxwell's Equations with DEC}
\label{sc2}
As mentioned above, the FDTD method can also be understood with DEC by viewing the average field values as a series of finite integrals. Just as in the Yee grid, it has been shown that the Maxwell's equations can also be expressed with DEC based on a simplicial mesh \cite{desbrun2005,desbrun2008,stern2015}. In this section, after a brief introduction to the framework of DEC, the governing equations for $3$-D and $2$-D problems in frequency domain are both formulated. 

\subsection{DEC approach to three-dimensional problems}
\begin{figure}[htbp]
	\centering
	\includegraphics[width = \textwidth]{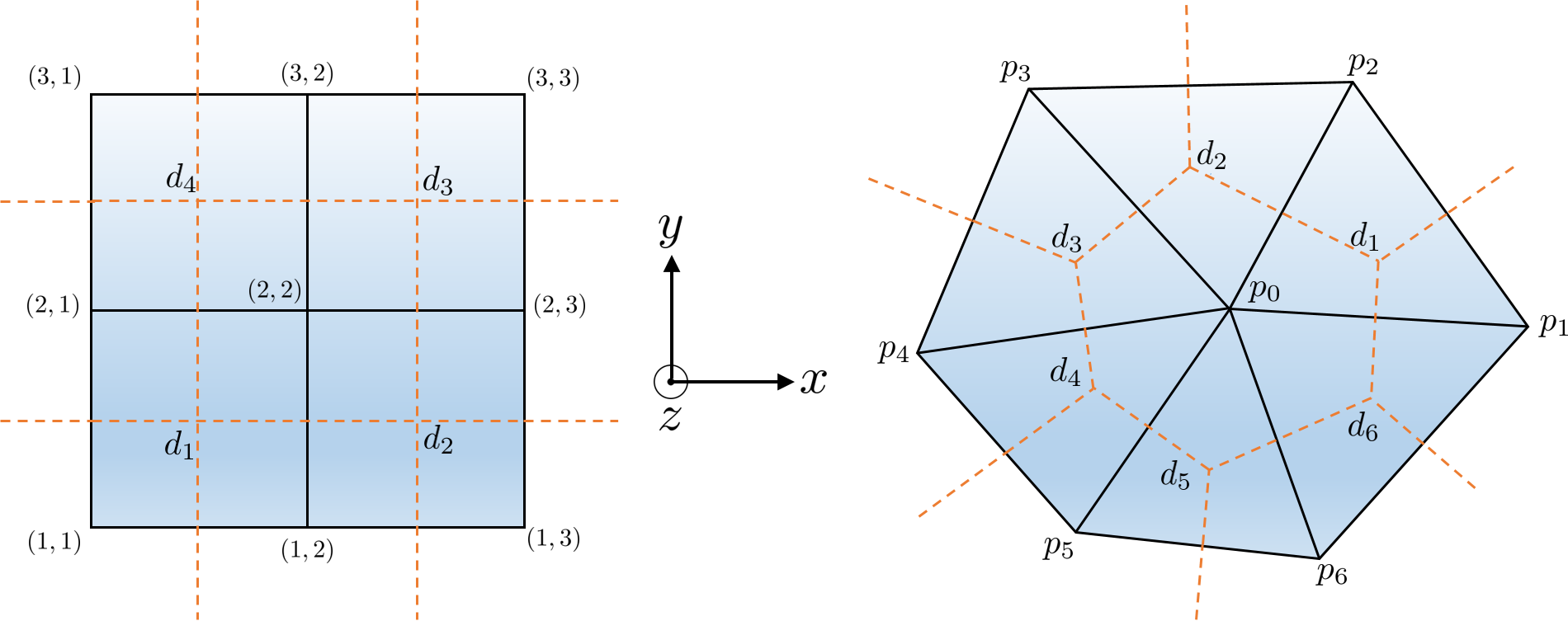}
	\caption{Left is the regular Yee grid on $x$-$y$ plane. Right is an example of a general triangular mesh (primal mesh) with its dual mesh (dashed) constructed by connecting nearest circumcenters. }
	\label{fig:y2d}
\end{figure}

As mentioned above, DEC is based on a simplicial mesh. Mathematically, $k$-simplex refers to the convex hull of $(k+1)$ vertices\footnote{The convex hull of a finite point set $S = \{x_i\}_{i=1}^N \subseteq \mathbb{R}^d$ refers to a region defined as Conv$(S) = \left\{\sum_{i=1}^N c_i x_i \bigg| c_i\ge0, \forall i; \sum_{i=1}^N c_i = 1 \right\}$.} in $k$-D space, e.g.\ points in $0$-D, lines in $1$-D, triangles in $2$-D and tetrahedrons in $3$-D. Figure \ref{fig:y2d} shows the comparison between a regular Yee grid and a general simplicial mesh in $2$-D. In DEC, the circumcenters are adopted to form the dual mesh. Therefore, a dual edge, denoted with $\c L_i$, is orthogonal to its related primal edge $l_i$, e.g.\ $\overline{d_1 d_2}$ and $\overline{p_0 p_1}$\footnote{Here, $\overline{v_1 v_2}$ is adopted to denote the line segment between vertices $v_1$ and $v_2$.} in Figure \ref{fig:y2d}. The dual face enclosed by $6$ dual edges and centered at $p_0$ is denoted with $\c A_{p_0}$. 

\begin{figure}[H]
	\begin{subfigure}{.32\textwidth}
		\centering
		\includegraphics[width = \textwidth]{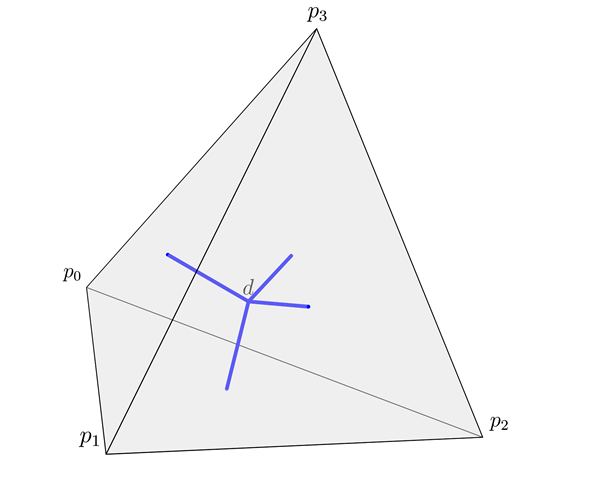}
		\caption{}
		\label{fig:d1_3D_eg}
	\end{subfigure}
	\begin{subfigure}{.32\textwidth}
		\centering
		\includegraphics[width = \textwidth]{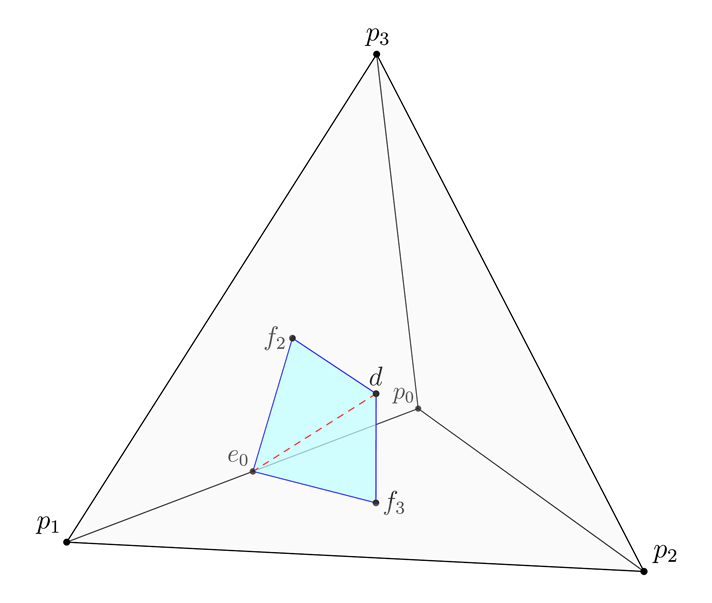}
		\caption{}
		\label{fig:d2_3D_eg}
	\end{subfigure}
	\begin{subfigure}{.32\textwidth}
		\centering
		\includegraphics[width = \textwidth]{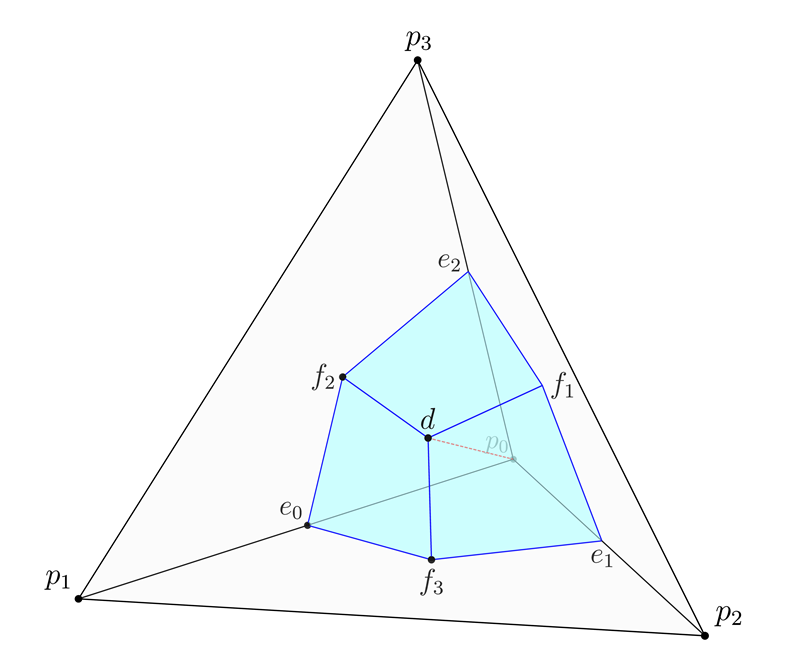}
		\caption{}
		\label{fig:d3_3D_eg}
	\end{subfigure}
	\caption{Dual edge, face and volume element in a single tetrahedron. The blue lines in (a) are 4 dual edges associating with 4 faces; the shaded area in (b) is a dual face associated with primal edge $\overline{p_0 p_1}$; the shaded volume in (c) is a dual volume related to $p_0$.}
	\label{fig:dual_3D}
\end{figure}

In $3$-D, the dual mesh is constructed by connecting circumcenters of nearest tetrahedrons. The dual edge, face and volume element in each tetrahedron is illustrated in Figure \ref{fig:dual_3D}. By definition, the orthogonality between primal and dual elements are also satisfied.

% In FDTD, the discrete version of $\nabla \times \v E = - \partial \v B / \partial t$ in the bottom-left square of Figure \ref{fig:y2d}, which we use $A_{d_1}$ to denote, is written as
% \begin{equation}
% -\frac{B_{z, d_1}^{n+1/2} - B_{z, d_1}^{n-1/2}}{\Delta t} = -\frac{E_{x, (2, 1+1/2)}^{n} - E_{x, (1, 1+1/2)}^{n}}{\Delta y} + \frac{E_{y, (1+1/2, 2)}^{n} - E_{y, (1+1/2, 1)}^{n}}{\Delta x}.
% \end{equation}
% Multiply both sides with $\Delta x \Delta y$, 
% \begin{align}
% \text{LHS} &\approx -\frac{1}{\Delta t} \left( \int_{A_{d_1}} B_z^{n+1/2} dA - \int_{A_{d_1}} B_z^{n-1/2} dA \right) \\
% \text{RHS} &\approx - \left( \int_{(2,1)}^{(2,2)} E_x^{n} dl - \int_{(1,1)}^{(1,2)} E_x^{n} dl \right) + \left( \int_{(1,2)}^{(2,2)} E_y^{n} dl - \int_{(1,1)}^{(2,1)} E_y^{n} dl \right) \nonumber \\
% & = \oint_{A_{d_1}} \v E \cdot d\v l
% \end{align}

Exterior calculus is induced by the Stokes' theorem
\begin{equation}
\int_{D} d \omega = \oint_{\partial D} \omega
\end{equation}
where $D$ and $\partial D$ represent any $k$-D domain and its $(k-1)$-D boundary, and $\omega$ is a $(k-1)$-form which can be integrated over any $(k-1)$-D space. Exterior derivative operator $d$, which transforms $(k-1)$-form $\omega$ to a $k$-form, acting on $0$-forms, $1$-forms and $2$-forms corresponds to gradient, curl and divergence operation in $3$-D space of vector calculus, respectively. In fact, Stokes' theorem is a generalization of Newton's, Green's and Gauss' theorem. However, in practice, both domain $D$ and its boundary $\partial D$ are discretized. And DEC is based on the discretized version of this theorem. In the following, we take Faraday's law  $d E = i\omega B$ as an example to illustrate this. Suppose primal face element $A_i$ (triangle) is enclosed by three edges, then the Stokes' theorem on $A_i$ is written as
\begin{equation}
\label{eqn:3}
i\omega \int_{A_i} B =  \int_{A_i} d E = \oint_{\partial A_i} E = \sum_{l_j \in \partial A_i} d_{i,j}^{(1)} \int_{l_j} E
\end{equation}
where $d_{i,j}^{(1)} \in \{+1, -1\}$ depends on if the direction of $l_j$ coincides with the direction of $A_i$, which can be clockwise or anti-clockwise. The superscript $(1)$ means this relation is about edges.

Then we can introduce the discrete counterpart of differential forms, namely \textit{cochains}\footnote{Chain, on the other hand, is used to denote a linear combination of simplices. In practice, a $k$-chain refers to a discrete domain a $k$-cochain can be integrated on.}, as column vectors, e.g.\ $1$-cochain $\bm E = (E_1, \dots, E_{N_1})^T$ and $2$-cochain $\bm B = (B_1, \dots, B_{N_2})^T$. Noted that $N_1$ and $N_2$ are the number of primal edge and face elements. The elements of these two cochains are defined as
\begin{equation}
\label{eqn:4}
E_j \triangleq \int_{l_j} E =  \int_{l_j} \left[ \v{E}(\v r) \cdot \hat{l}_j \right] dl, \quad B_i \triangleq \int_{A_i} B = \int_{A_i} \left[ \v{B}(\v r) \cdot \hat{n}_i\right] dA
\end{equation}
where $\hat{l}_j$ is the unit vector along $l_j$, and $\hat{n}_i$ is the unit vector normal to $A_i$. It should be pointed out that although traditionally cochain $\bm E$ and $\bm B$ are defined on primal mesh and are called primal cochains, they can also be defined on edges and faces of dual mesh. Then the relations described in Equation (\ref{eqn:3}) can be represented as
\begin{equation}
\label{eqn:5}
i \omega \bm B = \mat{d}^{(1)} \bm E
\end{equation}
where $\mat{d}^{(1)}$, as an $N_2 \times N_1$ matrix, is the discrete version of exterior derivative or \textit{co-boundary} operator, and the superscript $(1)$ means that the operator only acts on primal $1$-cochains. The $(i,j)$ element of this matrix is defined as
\begin{equation}
\label{eqn:6}
\left[ \mat{d}^{(1)} \right]_{i,j}= d_{i, j}^{(1)} = \begin{cases} \pm 1 & \text{if $l_j$ is an edge of $A_i$,} \\ 0 &  \text{otherwise.} \end{cases}
\end{equation}

Similarly, we can also define other co-boundary operators $\mat{d}^{(2)}$ (an $N_3 \times N_2$ matrix) acting on primal $2$-cochains and $\mat{d}^{(0)}$ (an $N_1 \times N_0$ matrix) acting on primal $0$-cochains as
\begin{align}
\label{eqn:7}
\left[ \mat{d}^{(2)} \right]_{i,j} &= d_{i, j}^{(2)} = \begin{cases} \pm 1 & \text{if $A_j$ is a face of $V_i$,} \\ 0 &  \text{otherwise,} \end{cases} \\
\left[ \mat{d}^{(0)} \right]_{i,j} &= d_{i, j}^{(0)} = \begin{cases} \pm 1 & \text{if $p_j$ is a vertex of $l_i$,} \\ 0 &  \text{otherwise,} \end{cases}
\label{eqn:8}
\end{align}
where $A_j$ and $V_i$ are the $j$-th primal face and $i$-th primal volume (a tetrahedron), and $p_j$ is the $j$-th primal vertex. 

Therefore, since the integral form of the divergence relation $d B = 0$ can be represented as
\begin{equation}
\int_{V_i} d B = \oint_{\partial V_i} B = \sum_{A_j \in \partial V_i} d_{i,j}^{(2)} \int_{A_j} B = 0
\end{equation} 
where $d_{i,j}^{(2)} \in \{+1, -1\}$ depends on if the normal direction of each face $A_j$ is pointing outside the volume $V_i$. Then this divergence relation can be described as
\begin{equation}
\label{eqn:10}
\mat{d}^{(2)} \bm B = 0.
\end{equation}

Different from primal cochains $\bm E$ and $\bm B$, cochains $\bm D$, $\bm H$, $\bm J$, $\bm \rho$ are defined on dual mesh and are called dual cochains. They are defined as
\begin{align}
\bm D &= (D_1, \dots, D_{N_1})^T, & \bm H &= (H_1, \dots, H_{N_2})^T, \nonumber \\
\bm J &= (J_1, \dots, J_{N_1})^T, & \bm \rho &= (\rho_1, \dots, \rho_{N_3})^T,
\end{align}
where
\begin{align}
D_i &\triangleq \int_{\c{A}_i} (\v{D}(\v r) \cdot \hat{l}_i) dA,  & H_i &\triangleq \int_{\c{L}_i} (\v{H}(\v r) \cdot \hat{n}_i) dl, \nonumber \\
J_i &\triangleq \int_{\c{A}_i} (\v{J}(\v r) \cdot \hat{l}_i) dA,  & \rho_i &\triangleq \int_{\c V_i} \rho(\v r) dV,
\label{eqn:12}
\end{align}
where $\c L_i, \c A_i$, and $\c V_i$ are the $i$-th dual edge, face and volume element, as illustrated in Figure \ref{fig:dual_3D}. Noted that due to orthogonality between primal and dual elements, $\hat{l}_i$, the unit vector along $i$-th primal edge $l_i$, is also the normal vector of dual face $\c{A}_i$, and $\hat{n}_i$, the unit normal vector through $i$-th primal face $A_i$, is parallel to $\c L_i$. Also noted that the number of dual edges is the same with the number of primal faces $N_2$, and the number of dual faces is the same with the number of primal edges $N_1$. And this explains the length of these dual cochains.

To build relations between dual cochains as in (\ref{eqn:5}) and (\ref{eqn:10}), co-boundary operators $\mat{d}_{\text{dual}}^{(k)}$ on dual mesh needs to be introduced. In fact there is a relation between co-boundary operators on $n$-dimensional primal and dual mesh \cite{desbrun2005, desbrun2008}
\begin{equation}
\label{eqn:13}
\mat{d}_{\text{dual}}^{(n-k)} = (-1)^k (\mat{d}^{(k-1)})^T
\end{equation}
where superscript $T$ refers to transpose. Specifically in $3$-D,
\begin{equation}
\mat{d}_{\text{dual}}^{(1)} = (\mat{d}^{(1)})^T, \quad \mat{d}_{\text{dual}}^{(2)} = - (\mat{d}^{(0)})^T, \quad \mat{d}_{\text{dual}}^{(0)} = (\mat{d}^{(2)})^T.
\end{equation}

Therefore, we can represent the discrete version of $d H = -i\omega D + J$ and $d D = \rho$ as 
\begin{align}
\label{eqn:15}
& \mat{d}_{\text{dual}}^{(1)} \bm H = (\mat{d}^{(1)} )^T \bm H = -i\omega \bm D + \bm J, \\
& \mat{d}_{\text{dual}}^{(2)} \bm D = - (\mat{d}^{(0)})^T \bm D = \bm \rho.
\end{align}
Moreover, since $\mat{d}^{(1)} \mat{d}^{(0)} = 0$ and $\mat{d}^{(2)} \mat{d}^{(1)} = 0$ exactly holds for both primal and dual mesh, another $\mat{d}^{(2)}_{\text{dual}}$ left multiplying to (\ref{eqn:15}) leads to the charge continuity relation
\begin{equation}
\mat{d}_{\text{dual}}^{(2)} \bm J = - (\mat{d}^{(0)})^T \bm J = i \omega \bm \rho,
\end{equation}   
which is naturally satisfied. An intuitive reason is that a $(k-2)$ simplex always appears twice with different signs in the boundary of the boundary of a $k$-simplex \cite{desbrun2008}. In fact, the exact preservation of this relation will prevent the generation of spurious charge in electromagnetic simulation. Therefore, only Equations (\ref{eqn:5}) and (\ref{eqn:15}) are independent.

Moreover, the constitutive relation can be illustrated by two characterized Hodge star operators, $N_1 \times N_1$ matrix $\star^{(1)}_{\epsilon}$ and $N_2 \times N_2$ matrix $\star^{(2)}_{\mu^{-1}}$. The construction scheme will be shown in the next section. These two operators map primal cochains $\bm E$ and $\bm B$ to dual cochains $\bm D$ and $\bm H$ as
\begin{align}
\label{eqn:cr3} % constitutive relation 3D
\bm D = \epsilon_0 \star^{(1)}_{\epsilon} \bm E, \quad \bm H = \mu_0^{-1} \star^{(2)}_{\mu^{-1}} \bm B.
\end{align}

Then combining Equations (\ref{eqn:5}), (\ref{eqn:15}) and (\ref{eqn:cr3}), we can obtain the equation for primal cochain $\bm E$ as
\begin{align}
\label{eqn:e_3d}
\left[ (\mat{d}^{(1)})^T \star^{(2)}_{\mu^{-1}} \mat{d}^{(1)} \right] \bm E = k_0^2 \left[\star^{(1)}_{\epsilon} \right]\bm E + i\omega \bm J,
\end{align}
where $k_0 = \omega\sqrt{\mu_0\epsilon_0}$ is the vacuum wave number. Then Equation (\ref{eqn:e_3d}) can be used to solve eigen modes in sourceless $3$-D space or the excitation field by a given current source. 

It should be pointed out that if the diagonal elements of $\star^{(2)}_{\mu^{-1}}$ are all positive, the semi-positiveness of $(\mat{d}^{(1)})^T \star^{(2)}_{\mu^{-1}} \mat{d}^{(1)}$ is guaranteed. For example, with arbitrary $1$-cochain $\bm e$,
\begin{equation*}
\bm e^T \left[ (\mat{d}^{(1)})^T \star^{(2)}_{\mu^{-1}} \mat{d}^{(1)} \right] \bm e = \bm b^T \star^{(2)}_{\mu^{-1}} \bm b \geq 0
\end{equation*}
where $\bm b = \mat{d}^{(1)} \bm e$. But if there is a $0$-cochain $\bm \phi$, such that $\bm e =\mat{d}^{(0)} \bm \phi$, cochain $\bm b$ is an exact zero cochain. Therefore, $(\mat{d}^{(1)})^T \star^{(2)}_{\mu^{-1}} \mat{d}^{(1)}$ has a large null space, just like $\nabla\times\nabla\times$ operator.

\subsection{Two-dimensional case}
Two-dimensional mesh, assumed in $x$-$y$ plane, can be thought of a degenerated $3$-D layered triangular mesh (extended along $z$-direction), or hexahedral mesh. In this $3$-D mesh, there are edges both on $x$-$y$ plane and along $z$-direction, and there are faces both on $x$-$y$ plane and parallel to $z$-direction. After degeneration to a $2$-D mesh, those edges and faces on $x$-$y$ plane remain the same, while the edges along $z$-direction become points and the faces parallel to $z$-direction become edges. Therefore, previous $1$-cochains associated with edges along $z$-direction become $0$-cochains in $2$-D mesh, and $2$-cochains associated with faces parallel to $z$-direction become $1$-cochains. In $2$-D problems, the fields are separated into $z$-component and transverse component, and each component corresponds to one cochain. By convention, we still place cochains associated with field $\v{E}(\v r)$ and $\v{B}(\v r)$ on primal mesh, and cochains related with field $\v{D}(\v r)$ and $\v{H}(\v r)$ on dual mesh. More specifically, the corresponding relations between fields and cochains are listed below:  
\begin{enumerate}
	\item 
	$z$-component:
	\begin{enumerate}
		\item
		$E_z(\v r) \hat{z}$ $\Rightarrow$ primal $0$-cochains $\bm E_z = (E_{z,1},\dots,E_{z,N_0})^T$;
		\item
		$B_z(\v r) \hat{z}$ $\Rightarrow$ primal $2$-cochain $\bm B_z = (B_{z,1},\dots, B_{z,N_2})^T$;
		\item
		$D_z(\v r) \hat{z}$ $\Rightarrow$ dual $2$-cochain $\bm D_z = (E_{z,1},\dots,E_{z,N_0})^T$;
		\item
		$H_z(\v r) \hat{z}$ $\Rightarrow$ dual $0$-cochains $\bm H_z = (H_{z,1},\dots, H_{z,N_2})^T$.
	\end{enumerate}
	\item
	transverse component:
	\begin{enumerate}
		\item
		$\v{E}_s(\v r)$ $\Rightarrow$ primal $1$-cochain $\bm E_s = (E_{s,1}, \dots, E_{s,N_1})^T$;
		\item
		$\v{B}_s(\v r)$ $\Rightarrow$ primal $1$-cochain $\bm B_s = (B_{s,1}, \dots, B_{s,N_1})^T$;
		\item
		$\v{D}_s(\v r)$ $\Rightarrow$ dual $1$-cochain $\bm D_s = (D_{s,1}, \dots, D_{s,N_1})^T$;
		\item
		$\v{H}_s(\v r)$ $\Rightarrow$ dual $1$-cochain $\bm H_s = (H_{s,1}, \dots, H_{s,N_1})^T$.
	\end{enumerate}
\end{enumerate}

For $0$-cochain $\bm E_z$ or $\bm H_z$, each element is just the field value at corresponding primal or dual vertex. For $2$-cochain $\bm B_z$ or $\bm D_z$, each element is the field integral on one primal face $A_i$ or dual face $\c A_i$ 
\begin{align}
\label{eqn:bzi}
B_{z,i} &\triangleq \int_{A_i} B_z(\v r) dA, \\
\label{eqn:dzi}
D_{z,i} &\triangleq \int_{\c{A}_i} D_z(\v r) dA.
\end{align}
Also noted that, although both $\bm E_s$ and $\bm B_s$ are $1$-cochains on primal mesh, they are defined in different ways
\begin{align}
\label{eqn:esi}
E_{s,i} &\triangleq \int_{l_i} \left[ \v{E}_s(\v r) \cdot \hat{l}_i \right] dl, \\
\label{eqn:bsi}
B_{s,i} &\triangleq \int_{l_i} \left[ \v{B}_s (\v r) \cdot (\hat{z} \times \hat{l}_i) \right] dl.
\end{align}
This is because $E_{s,i}$ denotes the potential change along $l_i$, while $B_{s,i}$ is the magnetic flux through $l_i$. In fact, $\bm B_s$ represents a $2$-cochain mesh in $3$-D defined on faces parallel to $z$-direction.
Similarly, dual $1$-cochain $\bm D_s$ and $\bm H_s$ are also defined differently
\begin{align}
\label{eqn:dsi}
D_{s,i} &\triangleq \int_{\c{L}_i} \left[ \v{D}_s(\v r) \cdot \hat{l}_i \right] dl, \\
\label{eqn:hsi}
H_{s,i} &\triangleq \int_{\c{L}_i} \left[ \v{H}_s (\v r) \cdot (\hat{z} \times \hat{l}_i) \right] dl.
\end{align}

The operator $\mat{d}^{(1)}$ and $\mat{d}^{(0)}$ in $2$-D mesh is defined the same way as in Equation (\ref{eqn:6}) and (\ref{eqn:8}). For dual mesh, according to (\ref{eqn:13}), we can obtain relations
\begin{equation}
\mat{d}^{(0)}_{\text{dual}} = (\mat{d}^{(1)})^T, \quad \text{and} \quad \mat{d}^{(1)}_{\text{dual}} = -(\mat{d}^{(0)})^T. 
\end{equation}

It should be pointed out that with $2$-D gradient operator $\nabla_s = \hat{x} \frac{\partial}{\partial x} + \hat{y} \frac{\partial}{\partial y}$, there are two possible operations on a transverse vector field: curl $\nabla_s \times$ and divergence $\nabla_s \cdot$. They are both represented by $\mat{d}^{(1)}$ or $\mat{d}^{(1)}_{\text{dual}}$, because they both acts on primal or dual $1$-cochains. However, observation of the $2$-D mesh in Figure \ref{fig:y2d} shows that primal edge $l_j$ enclosing primal face $A_i$ clockwise contradicts with $\hat{z} \times \hat l_j$ pointing outside $A_i$, and dual edge $\c L_j$ (with direction $\hat{z} \times \hat l_j$) enclosing dual face $\c A_i$ clockwise coincide with $\hat l_j$ pointing outside $\c A_i$. Therefore, divergence operation $\nabla_s \cdot$ acting on primal $1$-cochains induces an extra negative sign to $\mat{d}^{(1)}$.

The constitutive relations in $2$-D involve two components. Assuming the permittivity and permeability matrices are 
\begin{equation}
\dyadg{\epsilon} = \begin{bmatrix} \dyadg{\epsilon}_s & 0 \\ 0 & \epsilon_{zz} \end{bmatrix}, \qquad
 \dyadg{\mu} = \begin{bmatrix} \dyadg{\mu}_s & 0 \\ 0 & \mu_{zz} \end{bmatrix},
\end{equation}
where $\dyadg{\epsilon}_s$ and $\dyadg{\mu}_s$ are $2\times2$ tensors functions and their components are along the transverse directions. For simplicity, we assume $\dyadg{\epsilon}_s = \epsilon_s (\v r) \mat I$ and $\dyadg{\mu}_s = \mu_s (\v r) \mat I$, and we use $\epsilon_z$ and $\mu_z$ to replace $\epsilon_{zz}$ and $\mu_{zz}$.
Then four Hodge star operators are needed, $\star^{(0)}_{\epsilon_z}$, $\star^{(1)}_{\epsilon_s}$, $\star^{(2)}_{\mu^{-1}_z}$, and $\star^{(1)}_{\mu^{-1}_s}$. The mapping relations are
\begin{align}
\bm D_z  &= \epsilon_0\star^{(0)}_{\epsilon_z} \bm E_z,& \ \bm D_s &= \epsilon_0\star^{(1)}_{\epsilon_s} \bm E_s, \\
\bm H_z  &=  \mu_0^{-1}\star^{(2)}_{\mu^{-1}_z} \bm B_z,& \ \bm H_s &= \mu_0^{-1}\star^{(1)}_{\mu^{-1}_s} \bm B_s.
\end{align}

For homogeneous waveguides or inhomogeneous ones with $k_z = 0$, e.g.\ $2$-D photonic crystals, TM and TE modes are decoupled and can be analyzed independently. The reduced wave equations of $E_z (\v r)$ or $H_z (\v r)$ for homogeneous waveguides can be written as:
\begin{align}
\nabla_s \cdot \nabla_s E_z (\v r) + k_s^2 E_z (\v r)  &= 0, \quad \text{for TM waves}, \\
\nabla_s \cdot \nabla_s H_z (\v r) + k_s^2 H_z (\v r)  &= 0, \quad \text{for TE waves}
\label{eqn:31}
\end{align}
where $k_s^2 = \omega^2 \mu\epsilon - k_z^2$.

For TM modes, $\nabla_s E_z (\v r)$ leads to a $1$-cochain on primal edges, which is $\mat{d}^{(0)} \bm E_z$. For the next operator $\nabla_s\cdot$ to function appropriately, a Hodge star operator $\star^{(1)}$ (no subscript due to homogeneity) needs to be inserted to transform $\mat{d}^{(0)} \bm E_z$ to it dual $1$-cochain.
Therefore, $\nabla_s \cdot \nabla_s$ can be replaced with $[ \mat{d}_{\text{dual}}^{(1)} \star^{(1)} \mat{d}^{(0)} ]$, or $- [ (\mat{d}^{(0)})^T \star^{(1)} \mat{d}^{(0)} ]$. Finally the TM mode governing equation of primal $0$-cochain $\bm E_z$ can be written with DEC as:
\begin{equation}
\label{eqn:tm}
\left[ (\mat{d}^{(0)})^T \star^{(1)} \mat{d}^{(0)} \right] \bm E_z - k_s^2 \left[ \star^{(0)} \right] \bm E_z = 0.
\end{equation}

For TE modes, the equation for dual $0$-cochain $\bm H_z$ can be similarly rewritten as:
\begin{equation}
\label{eqn:te}
\left[ \mat{d}^{(1)} (\star^{(1)})^{-1} (\mat{d}^{(1)})^T \right] \bm H_z - k_s^2 \left[ (\star^{(2)})^{-1} \right] \bm H_z = 0.
\end{equation}

For inhomogeneous waveguide with $k_z = 0$, Equation (\ref{eqn:tm}) and (\ref{eqn:te}) need to be modified accordingly.
\begin{align}
\label{eqn:tm_in}
\text{TM:} &\quad \left[ (\mat{d}^{(0)})^T \star^{(1)}_{\mu_s^{-1}} \mat{d}^{(0)} \right] \bm E_z - k_0^2 \left[ \star^{(0)}_{\epsilon_z} \right] \bm E_z = 0, \\
\label{eqn:te_in}
\text{TE:} &\quad \left[ \mat{d}^{(1)} (\star^{(1)}_{\epsilon_s})^{-1} (\mat{d}^{(1)})^T \right] \bm H_z - k_0^2 \left[ (\star^{(2)}_{\mu_s^{-1}})^{-1} \right] \bm H_z = 0
\end{align}
where $k_0^2 = \omega^2 \mu_0 \epsilon_0$.

For inhomogeneous waveguide with nonzero $k_z$, TE and TM modes are coupled to each other, and the resulting modes are hybrid. The governing equation of the transverse field $\v E_s(\v r)$ and $\v H_s(\v r)$ can be written as \cite{chew1995}:
\begin{equation}
\label{eqn:ine}
\dyadg{\mu}_s \cdot \hat{z} \times \nabla_s \times \mu_{z}^{-1} \nabla_s \times \v{E}_s - \hat{z} \times \nabla_s \epsilon_{z}^{-1} \nabla_s \cdot \dyadg{\epsilon}_s \cdot \v{E}_s  
- k_0^2 \dyadg{\mu}_s \cdot \hat{z} \times  \dyadg{\epsilon}_s \cdot \v{E}_s = 
-k_z^2 \hat{z} \times \v{E}_s,
\end{equation}
and
\begin{equation}
\label{eqn:inh}
\dyadg{\epsilon}_s \cdot \hat{z} \times \nabla_s \times \epsilon_{z}^{-1} \nabla_s \times \v{H}_s - \hat{z} \times \nabla_s \mu_{z}^{-1} \nabla_s \cdot \dyadg{\mu}_s \cdot \v{H}_s  
- k_0^2 \dyadg{\epsilon}_s \cdot \hat{z} \times  \dyadg{\mu}_s \cdot \v{H}_s = 
-k_z^2 \hat{z} \times \v{H}_s. 
\end{equation}

For example, we rewrite Equation (\ref{eqn:ine}) with DEC one by one. As mentioned, transverse field $\v{E}_s(\v r)$ is represented as a primal $1$-cochain $\bm E_s$.
And $\dyadg{\epsilon}_s$, $\dyadg{\mu}_s^{-1}$, $\epsilon_{z}$, and $\mu_{z}^{-1}$ can be replaced by four Hodge star operators. It should also be pointed out that although $\hat{z} \times $ rotates the vector field by $90$ degree, the corresponding cochain remains the same. 
For differential operators, in the first term of (\ref{eqn:ine}), first $\nabla_s \times$ acting on primal $1$-cochains $\bm E_s$ refers to $\mat{d}^{(1)}$, and results in a primal $2$-cochain. Then subsequent $\dyadg{\mu}_s \cdot \hat{z}\times \nabla_s \times \mu_z^{-1}$ acting on this primal $2$-cochain\footnote{For compactness, we only defined Hodge star operators mapping primal to dual cochains. If a reverse mapping is needed, we use the inverse of the corresponding Hodge star operator.} leads to $(\star_{\mu_s^{-1}}^{(1)})^{-1} \mat{d}^{(0)}_{\text{dual}}\star_{\mu_z^{-1}}^{(2)}$. In the second term of (\ref{eqn:ine}), $\nabla_s \cdot \dyadg{\epsilon}_s$ acting on $\bm E_s$ refers to $\mat{d}^{(1)}_{\text{dual}} \star^{(1)}_{\epsilon_s}$, and results in a dual $2$-cochain. Then the subsequent gradient operator $\nabla_s \epsilon_z^{-1}$ acting on this dual $2$-cochains  leads to $\mat{d}^{(0)} (\star_{\epsilon_z}^{(0)})^{-1}$. The third term of (\ref{eqn:ine}) only involves two Hodge star operators.
Therefore, (\ref{eqn:ine}) can be transformed to
\begin{equation}
\label{eqn:38}
\begin{split}
\left[ (\star_{\mu_s^{-1}}^{(1)})^{-1} (\mat{d}^{(1)})^T \star_{\mu_z^{-1}}^{(2)} \mat{d}^{(1)} \right] \bm E_s   + \left[ \mat{d}^{(0)}  (\star_{\epsilon_z}^{(0)} )^{-1} (\mat{d}^{(0)})^T \star_{\epsilon_s}^{(1)} \right] \bm E_s &\\
 - k_0^2 \left[ (\star_{\mu_s^{-1}}^{(1)})^{-1} \star_{\epsilon_s}^{(1)} \right] \bm E_s = - k_z^2 \bm E_s &.
\end{split}
\end{equation}

Therefore, Equations (\ref{eqn:tm}) to (\ref{eqn:te_in}) and Equation (\ref{eqn:38}) can be applied with certain boundary conditions to solve eigen mode problems for $2$-D structures.

\section{Hodge Star Operators}
\label{sc3}
Hodge star operators defined in Section 2 map a primal cochain to its corresponding dual cochain. It should be pointed out the ``orthogonal'' dual and corresponding constitutive relation was first proposed in \cite{bossavit2000}, then fully developed in the frame of DEC. From the definition for $E_i$ and $D_i$ in Equation (\ref{eqn:4}) and (\ref{eqn:12}), with locally constant field assumption, we can infer that there should be a one to one relation between them
\begin{equation}
\label{eqn:ed_con}
\int_{\c{A}_i} \left(\v{D}(\v r) \cdot \hat{l}_i \right) dA = \epsilon_i \int_{\c{A}_i} \left(\v{E}(\v r) \cdot \hat{l}_i \right) dA = (\star^{(1)}_{\epsilon})_i \int_{l_i} \left( \v{E}(\v r) \cdot \hat{l}_i \right) dl.
\end{equation}
Here, $(\star^{(1)}_{\epsilon})_i$ is the $i$-th diagonal matrix element of $\star^{(1)}_{\epsilon}$, and $\epsilon_i$ is the average permittivity. Then $(\star^{(1)}_{\epsilon})_i$ is defined to be the ratio of dual face element's area to primal edge element's length, and multiplied by $\epsilon_i$. Note that, from now on, we use \textit{volume} as a general term for length, area and volume in $1$-D, $2$-D and $3$-D, respectively. In this case, since the local value of $\epsilon$ needs to be included to obtain a \textit{characteristic} Hodge star, and this is why $\epsilon$ is in the subscript. If we adopt this definition, (\ref{eqn:ed_con}) can be written as
\begin{equation}
\int_{\c{A}_i} \left(\v{D} \cdot \hat{l}_i \right) dA = \frac{\epsilon_i |\c{A}_i|}{|l_i|} \int_{l_i} \left( \v{E} \cdot \hat{l}_i \right) dl.
\end{equation}
Here, $|\c{A}_i|$ and $|l_i|$ denote the volume of dual face $\c{A}_i$ and primal edge $l_i$. Similarly the relation between $H_i$ and $B_i$ defined in (\ref{eqn:4}) and (\ref{eqn:12}) reads
\begin{equation}
\int_{\c{L}_i} \left( \v{H} \cdot \hat{n}_i \right) dl = \frac{\mu_i^{-1} |\c{L}_i|}{|A_i|} \int_{A_i} \left( \v{B} \cdot \hat{n}_i \right) dA
\end{equation}
where $|\c{L}_i|$ and $|A_i|$ are the volume of dual edge $\c{L}_i$ and primal face $A_i$, and $\mu_i$ is the local value of $\mu(\v r)$. Then the Hodge star operator $\star^{(1)}_{\epsilon}$ and $\star^{(2)}_{\mu^{-1}}$ in $3$-D are constructed as below
\begin{equation}
\star^{(1)}_{\epsilon} = 
\begin{bmatrix} \frac{\epsilon_1 |\c{A}_1|}{|l_1|} & & \\ & \ddots & \\ & & \frac{\epsilon_{N_1} |\c{A}_{N_1}|}{|l_{N_1}|} \end{bmatrix}, \quad
\star^{(2)}_{\mu^{-1}} = 
\begin{bmatrix} \frac{\mu^{-1}_1 |\c{L}_1|}{|A_1|} & & \\ & \ddots & \\ & & \frac{\mu^{-1}_{N_1} |\c{L}_{N_1}|}{|A_{N_1}|} \end{bmatrix}.
\label{eqn:hd_mat}
\end{equation}

In a $2$-D case, from the definition in Equation (\ref{eqn:bzi})-(\ref{eqn:hsi}), there are four constitutive relations involved
\begin{align}
D_{z,i}& = (\star_{\epsilon_z})_i E_{z,i} = \epsilon_{z,i} |\c A_i| E_{z,i},& D_{s,i}& = (\star_{\epsilon_s})_i E_{s,i} = \frac{\epsilon_{s,i}|\c L_i|}{l_i} E_{s,i}, \\
H_{z,i}& = (\star_{\mu_z^{-1}})_i B_{z,i} = \frac{\mu_{z,i}^{-1}}{|\c A_i|} B_{z,i},& H_{s,i}& = (\star_{\mu_s^{-1}})_i B_{s,i} = \frac{\mu_{s,i}^{-1}|\c L_i|}{l_i} B_{s,i}
\end{align}
where $E_{z,i}$ and $H_{z,i}$ are the field value at $i$-th primal and dual vertex.

Noted that, in practice, $\epsilon(\v r)$ and $\mu(\v r)$ are assumed to be constant in one tetrahedron (triangle in $2$-D). However, since generally dual edge elements or dual face elements do not belong to a unique tetrahedron, a weighted average needs to be performed, which will be discussed next.
It should be pointed out that even with Delaunay triangulation, circumcenter may falls outside the tetrahedron (triangle in $2$-D). However, there are some problems for which Delaunay triangulation is not a good idea. In this volume calculation, we also put this special case into consideration and show that our scheme is suitable for a general triangular mesh (or tetrahedral mesh). We will start from a $2$-D case.

\subsection{Volume of dual cells in 2D} 
\begin{figure}[htbp]
	\centering
	\includegraphics[width = 0.6\textwidth]{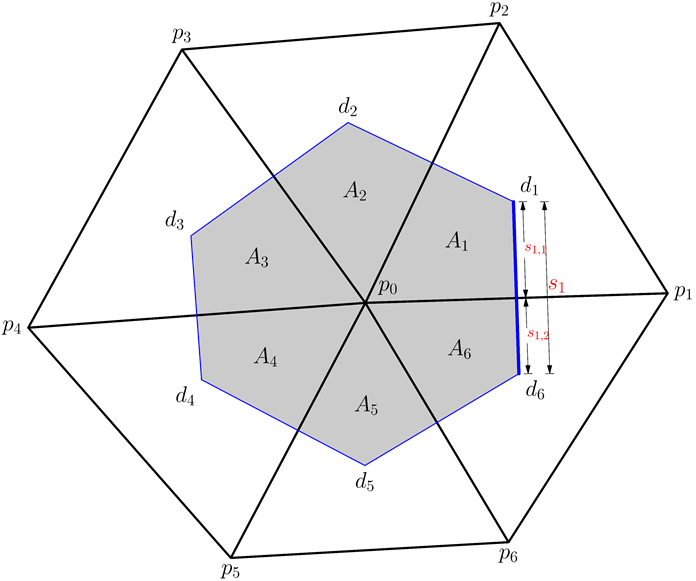}
	\caption{A dual $2$-cell $\c{A}_{p_0}$ (shaded in gray) formed by $6$ circumcenters. The bold blue edge $\overline{d_1 d_6}$ is associated with primal edge $\overline{p_0 p_1}$. }
	\label{fig:mtd_2D}
\end{figure}
Then in a simple $2$-D mesh, as shown in Figure \ref{fig:mtd_2D}, dual edge elements, such as $\c{L}_{d_1 d_6}$, are composed by two components from two neighboring triangles ($\bigtriangleup_{p_0,p_1,p_2}$ and $\bigtriangleup_{p_0, p_6, p_1}$), while dual face elements are composed by several parts from triangles sharing the same primal vertex. For example, dual face $\c{A}_{p_0}$ associated with $p_0$ is composed by 6 parts $A_1, \dots, A_6$.  Then we can obtain the length of dual edge $\c{L}_{d_1 d_6}$ by adding its two components, and the area of gray shaded $2$-cell by summing over all its six parts:
\begin{align}
\label{eqn:33}
|\c{L}_{d_1 d_6 }| & = s_1 = s_{1,1} + s_{1,2}, \\
\label{eqn:34}
|\c{A}_{p_0}| & = \sum_{i = 1}^6 |A_i| = \frac{1}{2} \sum_{i=1}^6 |\c L_i| l_i.
\end{align}
Here the second equality sign in (\ref{eqn:34}) is because that the dual face $\c A_{p_0}$ can also be decomposed into $6$ triangles, and each of them has one dual edge as bottom edge and $p_0$ as another vertex. In this way, the volume of dual edge and face elements can be calculated systematically.

With inhomogeneous material information, a weighted average needs to be performed to obtain effective dual volume. For example, if each region $A_i$ in Figure \ref{fig:mtd_2D} is associated with a distinct permittivity value $\epsilon_i$, then effective volume of $\c{L}^*_{d_1 d_6, \epsilon}$ and $\c{A}^*_{p_0, \epsilon}$ can be obtained as
\begin{align}
\label{eqn:35}
|\c{L}^*_{d_1 d_6, \epsilon}| & = \epsilon_1 s_{1,1} + \epsilon_2 s_{1,2}, \\
\label{eqn:36}
|\c{A}_{p_0, \epsilon}^*| & = \sum_{i = 1}^6 \epsilon_i |A_i|= \frac{1}{2} \sum_{i=1}^6 |\c L_{i, \epsilon}^*| l_i.
\end{align}

Therefore, Hodge star operators $\star^{(0)}_{\epsilon_z}$ and $\star^{(1)}_{\epsilon_s}$ can be constructed as follows
\begin{equation}
\star^{(0)}_{\epsilon_z} = 
\begin{bmatrix} |\c{A}^*_{1, \epsilon_z}| & & \\ 
& \ddots &  \\ 
& & |\c{A}^*_{N_0, \epsilon_z}| \end{bmatrix}, \quad
\star^{(1)}_{\epsilon_s} = 
\begin{bmatrix} \frac{|\c{L}^*_{1,\epsilon_s}|}{|l_1|} & & \\ 
& \ddots & \\ 
& & \frac{|\c{L}^*_{N_1,\epsilon_s}|}{|l_{N_1}|} 
\end{bmatrix}.
\end{equation}
Noted that, $\epsilon_z$ and $\epsilon_s$ can be chosen differently for anisotropic material.

Similarly, we can also obtain effective dual volume based on localized value of $\mu^{-1}$
\begin{align}
|\c{L}^*_{d_1d_6, \mu^{-1}}| & = (\mu_1)^{-1} s_{1,1} + (\mu_2)^{-1} s_{1,2}, \\
|A_{d_i, \mu^{-1}}^*| & = (\mu_i)^{-1} |A_{d_i}|, \quad i = 1, \dots, 6.
\end{align}
Here $A_{d_i}$ is the primal triangle with circumcenter $d_i$.

Then Hodge star operators $\star^{(2)}_{\mu_z^{-1}}$ and $\star^{(1)}_{\mu_s^{-1}}$ can be constructed as
\begin{align}
\star^{(2)}_{\mu_z^{-1}} = 
\begin{bmatrix} |A^*_{1, \mu_z^{-1}}| & & \\
& \ddots & \\ 
& & |A^*_{N_2, \mu_z^{-1}}| \end{bmatrix}, \
\star^{(1)}_{\mu_s^{-1}} = 
\begin{bmatrix} 
\frac{|\c{L}^*_{ 1,\mu_s^{-1}}| }{ |l_1| } & &\\ 
& \ddots & \\ 
& & \frac{ |\c{L}^*_{N_1,\mu_s^{-1}}| }{ |l_{N_1}| } 
\end{bmatrix}.
\end{align}
In the above, $\mu_z$ and $\mu_s$ can also be chosen differently.

If an inverse mapping is needed, such as $\bm D$ to $\bm E$ and $\bm H$ to $\bm B$, we can just use the direct inverse of these Hodge star operators since they are all diagonal.

\subsubsection{Special case: dual vertex is outside the triangle}
\begin{figure}[H]
\begin{subfigure}{0.5\textwidth}
	\centering
	\includegraphics[width = \textwidth]{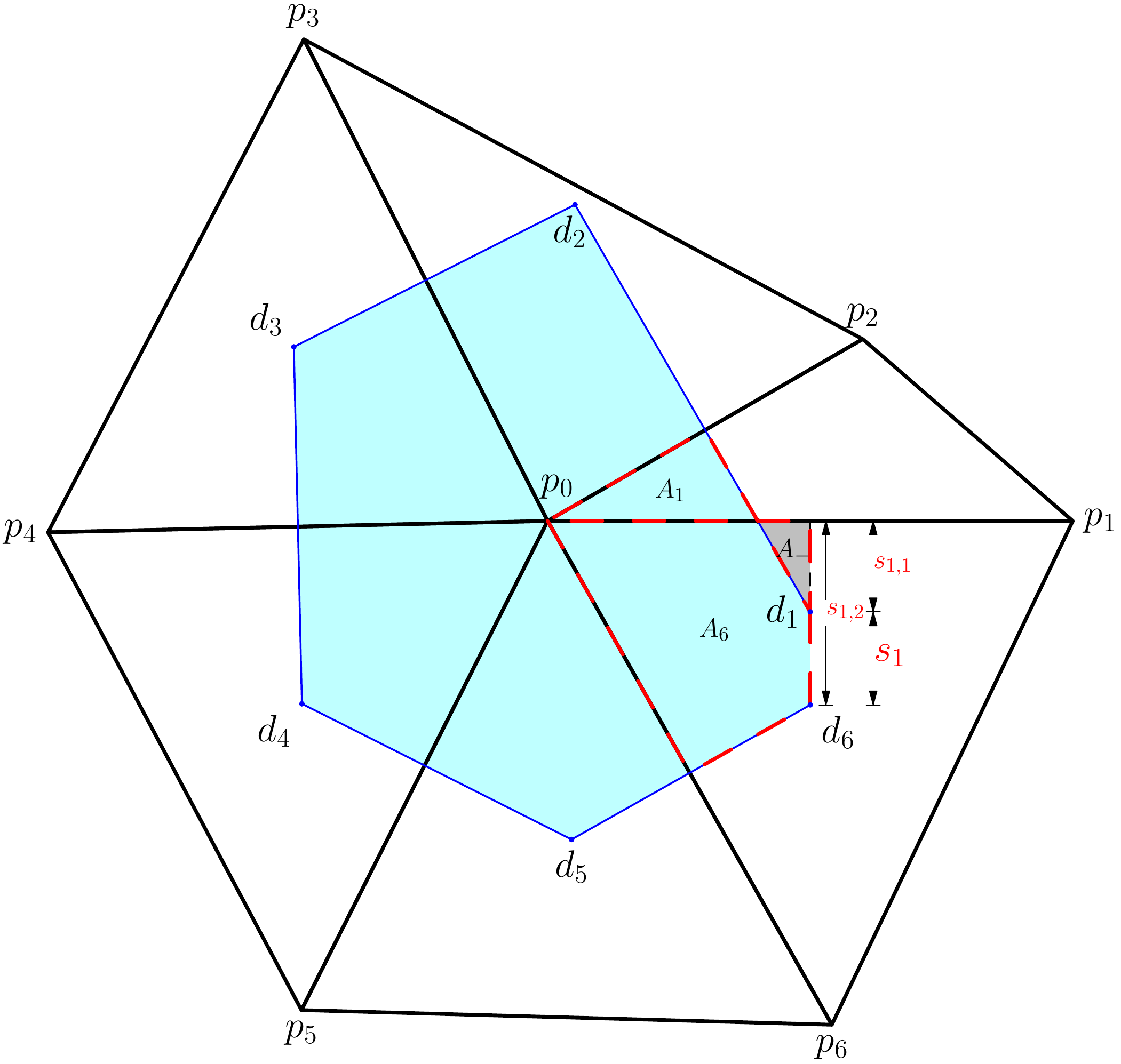}
	\caption{}
	\label{fig:mt_out1}
\end{subfigure}
\begin{subfigure}{0.5\textwidth}
	\centering
	\includegraphics[width = 0.9\textwidth]{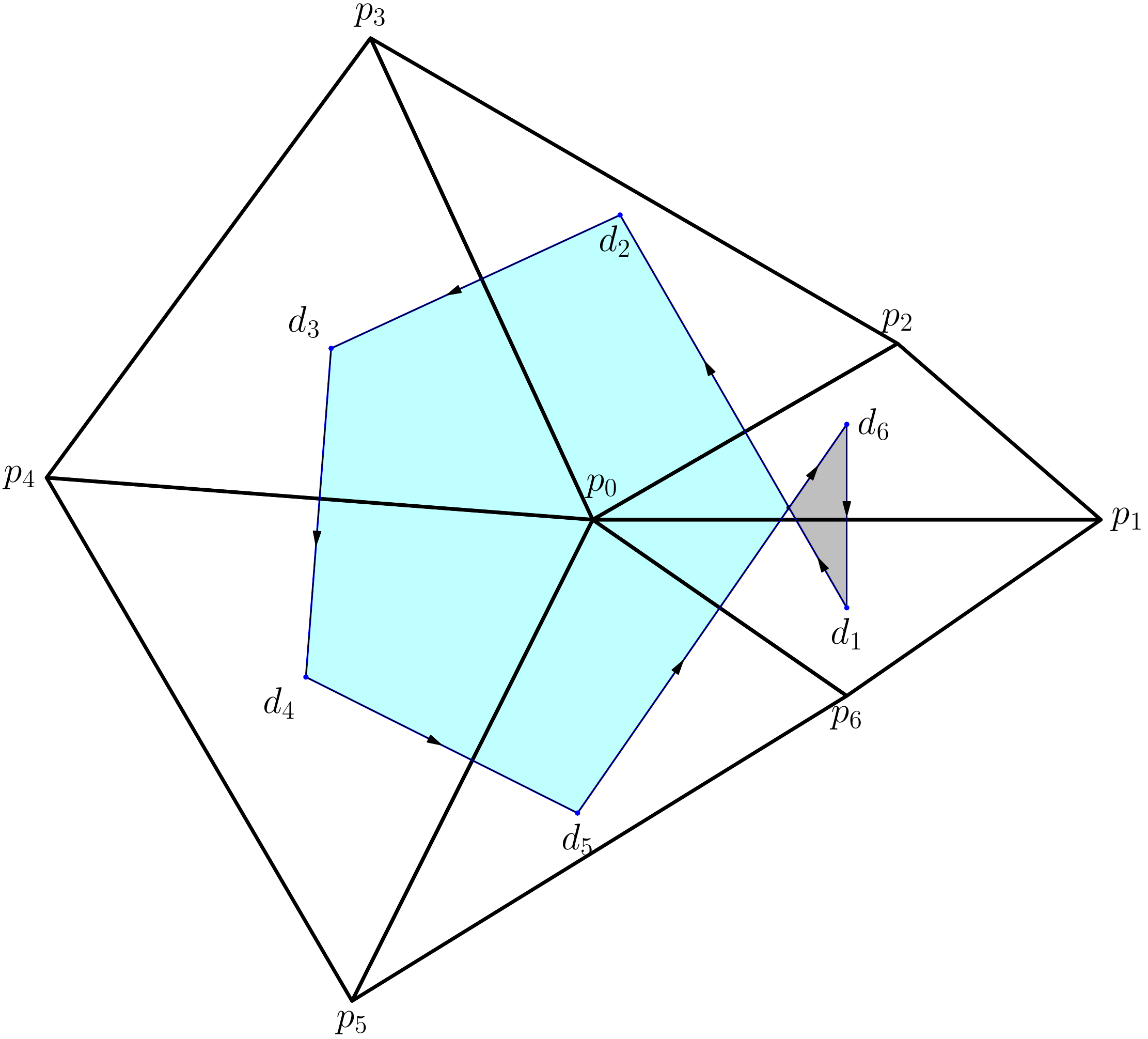}
	\caption{}
	\label{fig:mt_out2}
\end{subfigure}
\caption{(a) Circumcenter $d_1$ is outside its associated triangle. $A_1, A_6, A_-$ denote the area enclosed by red dashed line, and $s_1$ is the length for dual edge $\overline{d_6 d_1}$; (b) An example of non-Delaunay triangulation. Circumcenters $d_1$ and $d_6$ are both outside their associated triangles. The cyan region means positive region, while the gray one represents a negative region.}
\label{fig:mt_out}
\end{figure}
In Figure \ref{fig:mt_out1}, a special mesh with one circumcenter ($d_1$) falling outside the associated triangle is shown, which is frequently encountered even with Delaunay triangulation. Then in (\ref{eqn:33}), $s_{1,1}$ is negative, while the calculation for $\c{A}_{p_0}$ based on the second equality in (\ref{eqn:34}) stays unchanged. It should be noted that since $s_1$, the volume of dual edge $\overline{d_6 d_1}$, is still positive, this mesh is still Delaunay triangulation ($p_2$ is outside the circumcircle of $\bigtriangleup_{p_0, p_1, p_6}.$).

The scheme described by Equation (\ref{eqn:34}) also works for an even more ``twisted'' case, a non-Delaunay triangulation, shown in Figure \ref{fig:mt_out2}. In fact, the total area of the dual face element can be found to be the difference between left cyan region and right gray region. From the arrow directions, we can see that the left cyan region has an anti-clockwise orientation, while the right gray region has a clockwise orientation. This means that they are \textit{signed} volume. Therefore, we need to subtract the clockwise one to obtain a corrected dual volume for this dual face.

\subsection{Volume of dual cells in 3D}
\begin{figure}[H]
\begin{subfigure}{0.5\textwidth}
	\centering
	\includegraphics[width = 0.95\textwidth]{d2_3D.png}
	\caption{}
	\label{fig:d2_3D}
\end{subfigure}
\begin{subfigure}{0.5\textwidth}
	\centering
	\includegraphics[width = 0.8\textwidth]{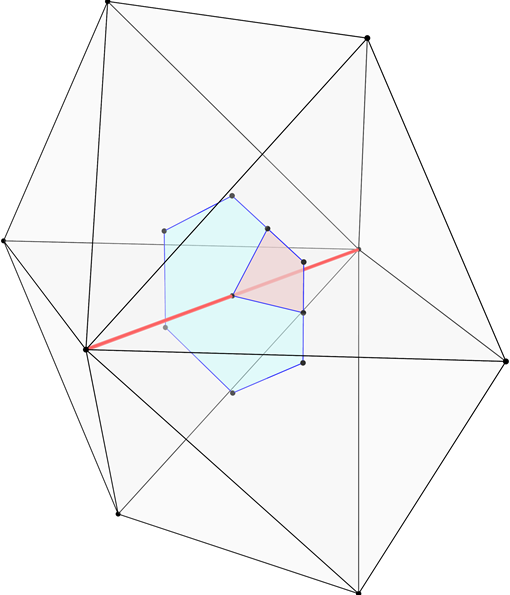}
	\caption{}
	\label{fig:mtet2}
\end{subfigure}
\caption{In (a), a dual face confined in a single tetrahedron is shown. Here, $d$ is the circumcenter of the tetrahedron, $f_2$, $f_3$ are circumcenters of corresponding triangles, and $e_0$ is the middle point of edge $\overline{p_0 p_1}$. (b) shows dual face associated with the red line in multiple tetrahedrons. The red shaded area is dual face confined to a single tetrahedron as in (a).}
\label{fig:d2_tet}
\end{figure}
The calculation for the volume of dual elements is far more complicated in $3$-D. We can calculate the volume of dual faces one tetrahedron by one tetrahedron, shown in Figure \ref{fig:d2_tet}. For example, the area for the shaded dual face in Figure \ref{fig:d2_3D} can be obtained as:
\begin{equation}
|\c{A}_{\overline{p_0 p_1}}| = |\bigtriangleup_{d, f_2, e_0}| + |\bigtriangleup_{d, e_0, f_3}| = \frac{1}{2} |\c{L}_{d f_2}| \cdot |\c{L}_{f_2 e_0}| + \frac{1}{2} |\c{L}_{d f_3}| \cdot |\c{L}_{f_3 e_0}|.
\label{equ:d2_vol}
\end{equation}
Here the subscript $\overline{p_0 p_1}$ is used because this dual face is associated with primal edge $\overline{p_0 p_1}$. Then the total volume for dual edge and face elements can be obtained by adding the values in a single tetrahedron appropriately. For example, as shown in Figure \ref{fig:mtet2}, the total area of shaded dual $2$-cell is composed by 6 components from 6 tetrahedrons. 

\begin{figure}[H]
	\centering
	\includegraphics[width = 0.4\textwidth]{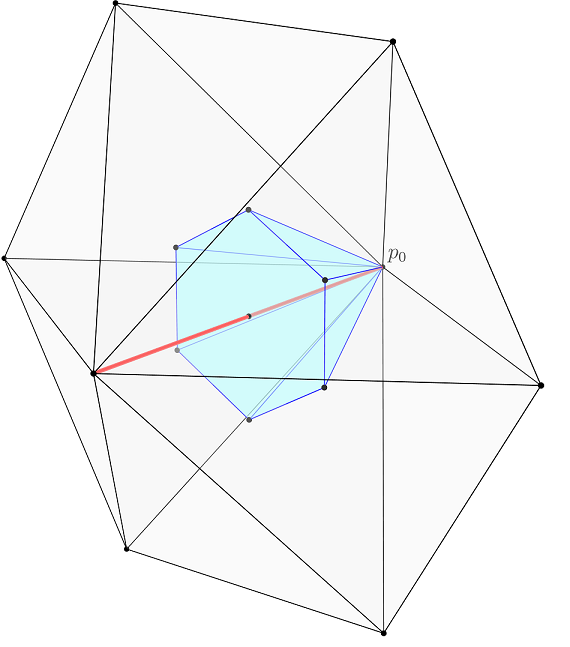}
	\caption{This is one component of dual polyhedron associated with vertex $p_0$. The number of components depends on how many dual faces this polyhedron has.}
	\label{fig:mtet3}
\end{figure}
As long as the volume of dual faces are obtained, the volume for dual $3$-cells can be obtained simply. Figure \ref{fig:mtet3} shows one component of the dual $3$-cell centered at vertex $p_0$. We can see that each of this component is a generalized cone associated with one dual face, or primal edge (the red line segment). Then the total volume for this dual $3$-cell $\c{V}_{p_0}$ can be obtained by summing over all its cone components as:
\begin{equation}
\label{eqn:43}
|\c{V}_{p_0}| = \sum_{\text{all its faces } \c{A}_i} \frac{1}{6} |\c{A}_i| \times |l_i|, \qquad l_i \text{ is the length of primal edge.}
\end{equation}
Here we have applied the volume formula of a cone shape structure: $V = \frac{1}{3}S \cdot h$, with bottom surface area $|\c{A}_i|$ and height $\frac{1}{2}|l_i|$. 

Incorporating the material information, like localized values of $\epsilon$ and $\mu^{-1}$, the effective volume can be obtained as
\begin{align}
|\c{L}^*_{\mu^{-1}}| & = \sum_{\text{components in neighboring tetrahedrons}} \mu^{-1}_i |\c{L}_i|, \\
|\c{A}^*_{\epsilon}| & = \sum_{\c{A}_i \text{ in tetrahedrons sharing the same edge}} \epsilon_i |\c{A}_i|, \\
|\c{V}^*_{\epsilon}| & = \sum_{\text{all faces of this dual $3$-cell }} \frac{1}{6} |\c{A}_{i,\epsilon}^*| \times |l_i|.
\end{align}

Therefore, the Hodge star operators $\star^{(1)}_{\epsilon}$ and $\star^{(2)}_{\mu^{-1}}$ can be constructed from Equation (\ref{eqn:hd_mat}).

\subsubsection{Special case: dual vertex is outside the tetrahedron}
\begin{figure}[H]
\begin{subfigure}{0.5\textwidth}
	\centering
	\includegraphics[width = 0.75\textwidth]{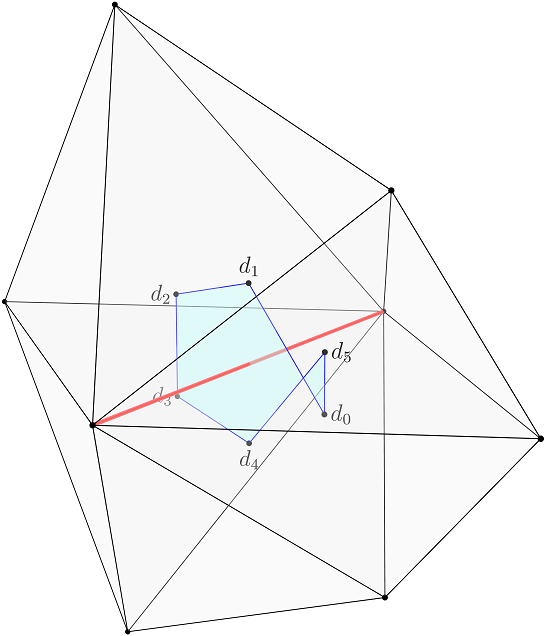}
	\caption{}
	\label{fig:mtet_out1}
\end{subfigure}
\begin{subfigure}{0.5\textwidth}
	\centering
	\includegraphics[width = 0.9\textwidth]{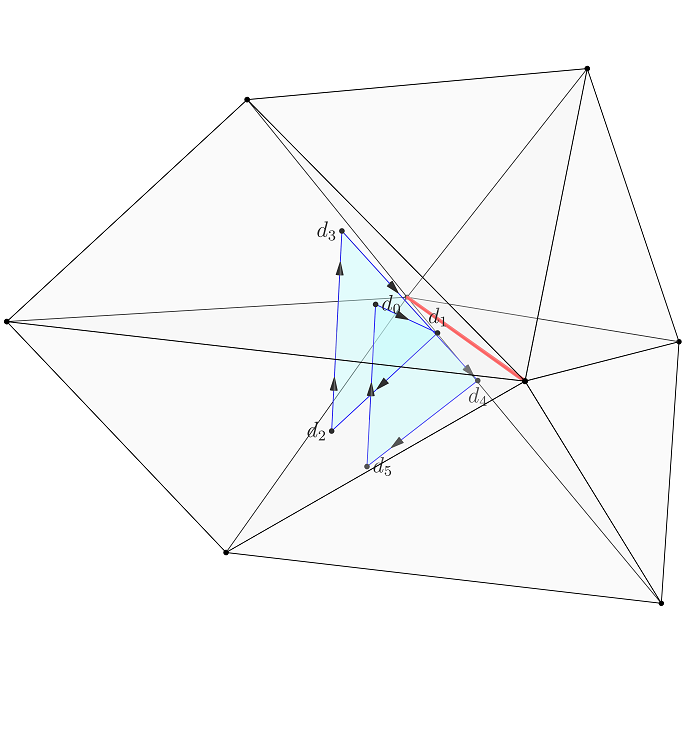}
	\caption{}
	\label{fig:mtet_out2}
\end{subfigure}
\caption{$d_0 \dots d_5$ are dual vertices. Cyan shaded area is the dual $2$-cell associated with the red edge. In (a), $d_0$  and $d_5$ are outside their associated tetrahedrons, while in (b), except for $d_1$, all dual vertices are outside.}
\label{fig:mtet_out}
\end{figure}

As shown in Figure \ref{fig:mtet_out}, since some dual vertices are outside their associated tetrahedrons, so the dual $2$-cell seems to be very ``twisted''. To obtain the corrected volume of dual cells for this special case, extra negative signs need to be assigned to some dual edges as we did in $2$-D,

Unlike in $2$-D, the complex geometry of dual face shown in Figure \ref{fig:mtet_out2} is originated from two reasons:
dual vertices fall outside tetrahedrons and
circumcenters of primal faces may also fall outside. 
This is why the red primal edge does not even go through its dual face. Two signs, instead of just one, are needed to take these two factors into account simultaneously. For example, as in Figure \ref{fig:d2_3D}, the area of triangle $\bigtriangleup_{d, f_2, e_0}$, $|\bigtriangleup_{d, f_2, e_0}|$, is written as:
\begin{equation}
|\bigtriangleup_{d, f_2, e_0}| = \frac{1}{2} |\c{L}_{d f_2}| \cdot |\c{L}_{f_2 e_0}|.
\end{equation}
But if $d$ is outside the tetrahedron, or $f_2$ is outside the triangle $\bigtriangleup_{p_0, p_1, p_3}$, extra signs need to be multiplied:
\begin{equation}
\text{multiplier} = \begin{cases} 
+1& \quad \text{if both $d$ and $f_2$ are inside,} \\  
-1& \quad \text{if only $d$ is outside,} \\ 
-1& \quad \text{if only $f_2$ is outside,} \\
+1& \quad \text{if both $d$ and $f_2$ are outside.} 
\end{cases}
\end{equation}
Then the following calculation for the volume of dual $3$-cells in Equation (\ref{eqn:43}) stays the same.

\section{Boundary Conditions}
\label{sc4}
In this section, we will show how to implement various of boundary conditions. The direct reason for a boundary condition is that the dual mesh is incomplete and truncated by primal mesh at the boundary, as illustrated in Figure \ref{fig:bc2}. In fact, PEC, PMC, and first-order ABC in time domain analysis have been investigated in \cite{rabina2014thesis}. We will show a different formulation for these boundary conditions in frequency domain, and in addition, we will also formulate second-order ABC and periodic boundary condition. For illustration, we use Equation (\ref{eqn:tm}) for $2$-D and Equation (\ref{eqn:e_3d}) for $3$-D. In fact, there is already one boundary condition embedded in $(\mat{d}^{(0)})^T$ for (\ref{eqn:tm}) and $(\mat{d}^{(1)})^T$ for (\ref{eqn:e_3d}). We first examine which kind this default boundary condition is.

\subsection{Default boundary condition: PMC}

\begin{figure}[htbp]
\centering
\includegraphics[width = 0.9\textwidth]{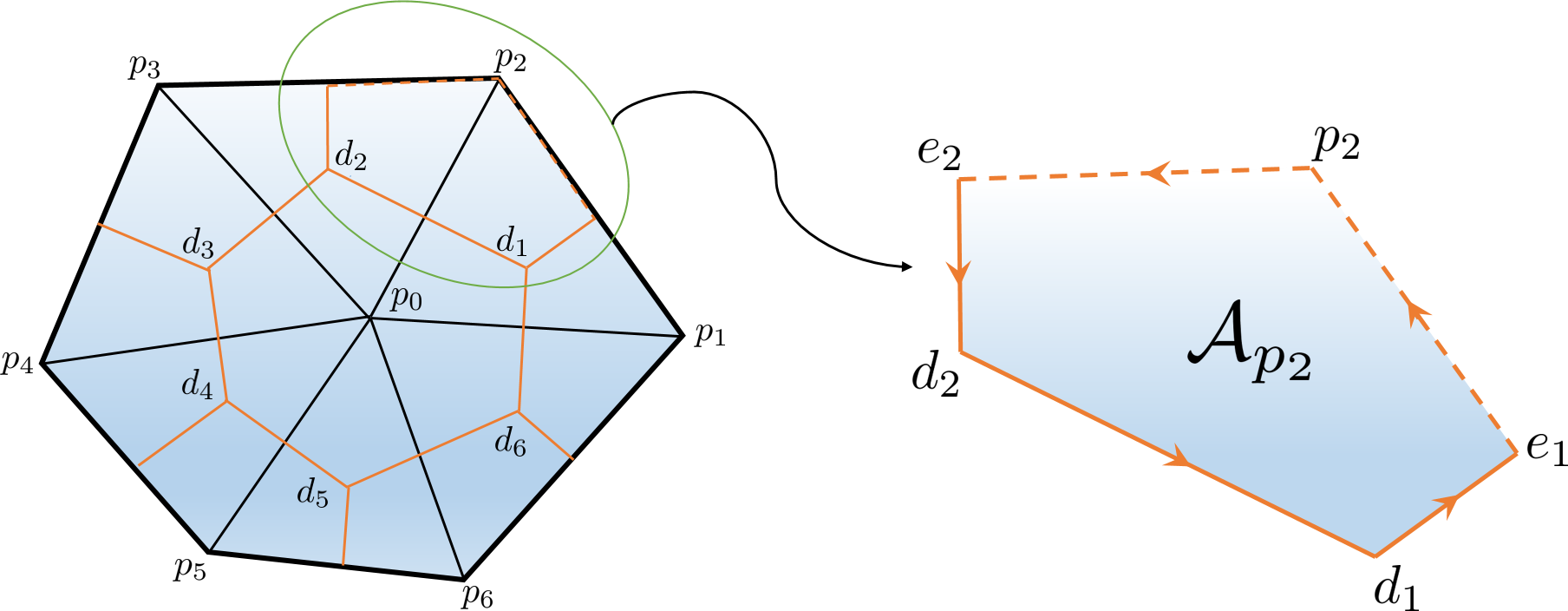}
\caption{Left is a primal triangular mesh with its dual mesh induced by $6$ circumcenters, $d_1, \dots, d_6$. The boundary is denoted by bold edges. Right is dual face $\c A_{p_2}$ with dashed edges on the boundary. Here, $e_1$ and $e_2$ are middle points of each edge. }
\label{fig:bc2}
\end{figure}
From (\ref{eqn:tm}), $\left[ \star^{(0)} \mat{d}^{(0)} \right] \bm E_z$ generates a dual $1$-cochain, and we denote it with $\bm H_s$. Then $\left[ (\mat{d}^{(0)})^T \right] \bm H_s$ leads to a dual $2$-cochain, which we represent it with $\bm D_z$. Next we use $D_{z,p_j}$ and $H_{s, \overline{v_i v_j}}$ to denote cochains' value on dual face $\c{A}_{p_j}$ and its surrounding dual edge $\overline{v_i v_j}$ ($v_i, v_j$ are two vertices) in Figure \ref{fig:bc2}. Without discretization, the relation between field $D_z(\v r) \hat{z}$ and $\v H_s(\v r)$ is 
\begin{equation*}
D_z(\v r) \hat{z} = \nabla_s \times \v H_s(\v r), \ \text{or} \ D_z(\v r) = \nabla_s \cdot \v H_s(\v r).
\end{equation*}
Then we can conclude the relation below from the integral form of this relation (Ampere's law):
\begin{align}
D_{z,p_2} &= \int_{\c A_{p_2}} D_z(\v r) dA = \oint_{\partial \c A_{p_2}} \v H_s(\v r) \cdot d\v l \nonumber \\
&= H_{s, \overline{e_2 d_2}} + H_{s, \overline{d_2 d_1}} + H_{s, \overline{d_1 e_1}} + H_{s, \overline{e_1 p_2}} + H_{s, \overline{p_2 e_2}}.
\label{eqn:49}
\end{align}
Here, $\partial \c A_{p_2}$ denotes the edges of dual face $\c A_{p_2}$. However, with discrete relation $\bm D_z = (\mat{d}^{(0)})^T \bm H_s$, only the first three terms in Equation (\ref{eqn:49}) are included. The last two terms $H_{s, \overline{e_1 p_2}}$, and $H_{s, \overline{p_2 e_2}}$ need to be determined by certain boundary conditions. Therefore, without extra boundary condition implemented, the last two terms of (\ref{eqn:49}) are set to be zero implicitly. And this is the perfect magnetic conductor (PMC) boundary condition with zero tangential magnetic field. For a $3$-D equation (\ref{eqn:e_3d}), this is also true. In the language of partial differential equation (PDE), this is the homogeneous natural (Neumann) boundary condition. 

\subsection{PEC and Dirichlet boundary condition}
Perfect electric conductor (PEC) boundary condition is a little different, and it refers to essential boundary condition in PDE. For TM modes, $\bm E_z$ defined on the boundary vertices all have zero value. Next we use subscript $\text{I}$ to denote field or value strictly inside the boundary, and subscript $B$ to denote field or value on the boundary. To consider the vertices strictly inside only and keep the constructed derivative operator, we can use a projection operator $\mat{P}_{E_z}$. Suppose there are $N_{0,\text{I}}$ vertices strictly inside out of total $N_0$ vertices. Then, $\mat{P}_{E_z}$ is a $N_0 \times N_{0, \text{I}}$ matrix, and it can insert zero boundary values to map a $\bm E_{z,\text{I}}$ cochain to the total $\bm E_z = \{ \bm E_{z,\text{I}}; \bm E_{z,\text{B}} \}$ cochain.
\begin{equation}
\label{eqn:50}
\mat{P}_{E_z} = \left(
\begin{array}{c}
\mat{I} \\
0 
\end{array} \right), \quad
\bm E_z = \mat{P}_{E_z} \cdot \bm E_{z,\text{I}}
\end{equation}
where $\mat{I}$ is $N_{0, \text{I}} \times N_{0, \text{I}}$ identity matrix.
Then (\ref{eqn:tm}) with PEC boundary condition should be adjusted as:
\begin{equation}
\label{eqn:63}
\left[\mat{P}_{E_z}\right]^T \left[(\mat{d}^{(0)})^T \star^{(1)} \mat{d}^{(0)} \right] \left[\mat{P}_{E_z} \right] \bm E_{z,\text{I}} = k_s^2 \left[ \mat{P}_{E_z}\right]^T  \left[ \star^{(0)}\right] \left[ \mat{P}_{E_z} \right] \bm E_{z,\text{I}}.
\end{equation}

For a Dirichlet boundary condition, e.g.\ in scattering problem, $E_{z, \text{scat}} = -E_{z, \text{inc}}$ must be satisfied for scattered field such that $E_{z, \text{total}} = 0$ on a PEC boundary. Therefore, by using operator $\mat{P}_{E_z}$, the scattered field cochain is represented as
\begin{equation}
\bm E_{z, \text{scat}} = \mat{P}_{E_z} \cdot \bm E_{z,\text{scat}, \text{I}} - \begin{pmatrix} 0 \\  \bm E_{z, \text{inc}, \text{B}} \end{pmatrix} = \begin{pmatrix} \bm E_{z,\text{scat}, \text{I}} \\ -\bm E_{z, \text{inc}, \text{B}}  \end{pmatrix}
\end{equation}
Then Equation (\ref{eqn:63}) for a $2$-D scattering problem is adjusted as
\begin{equation}
\label{eqn:tm_scat}
\begin{split}
\left[\mat{P}_{E_z}\right]^T & \left[(\mat{d}^{(0)})^T \star^{(1)} \mat{d}^{(0)} -  k_0^2 \star^{(0)} \right]  \left[\mat{P}_{E_z} \right] \bm E_{z,\text{I}} \\
& = \left[\mat{P}_{E_z}\right]^T \left[(\mat{d}^{(0)})^T \star^{(1)} \mat{d}^{(0)} \right] \begin{pmatrix} 0 \\ \bm E_{z, \text{inc}, \text{B}} \end{pmatrix}.
\end{split}
\end{equation}

For Equation (\ref{eqn:38}), PEC boundary condition can be implemented similarly by introducing projection operator $\mat{P}_{E_s}$. Similarly, $\mat{P}_{E_s}$ is a $N_1 \times N_{1, I}$ matrix with $N_1$ and $N_{1, I}$ representing the number of all edges and edges strictly inside. Then with $\bm E = \{\bm E_{s, I}; \bm E_{s, B}\}$, (\ref{eqn:38}) with PEC is written as
\begin{equation}
\label{eqn:38_pec}
\begin{split}
\left[\mat P_{E_s} \right]^T \left[ (\star_{\mu_s^{-1}}^{(1)})^{-1} (\mat{d}^{(1)})^T \star_{\mu_z^{-1}}^{(2)} \mat{d}^{(1)}  +  \mat{d}^{(0)}  (\star_{\epsilon_z}^{(0)} )^{-1} (\mat{d}^{(0)})^T \star_{\epsilon_s}^{(1)} \right] \left[\mat P_{E_s} \right] \bm E_{s,I} &\\
 - k_0^2 \left[\mat P_{E_s} \right]^T \left[ (\star_{\mu_s^{-1}}^{(1)})^{-1} \star_{\epsilon_s}^{(1)} \right]\left[\mat P_{E_s} \right] \bm E_{s,I} = - k_z^2 \left[\mat P_{E_s}^T\mat P_{E_s} \right]\bm E_{s,I} &,
\end{split}
\end{equation}
with
\begin{equation}
\mat{P}_{E_s} = \left(
\begin{array}{c}
\mat{I} \\
0 
\end{array} \right), \quad
\bm E_s = \mat{P}_{E_s} \cdot \bm E_{s,I}
\end{equation}
where $\mat{I}$ is $N_{1, I} \times N_{1, I}$ identity matrix.

However, this is not the only way to implement PEC boundary condition. Inspired by the embedded PMC boundary condition discussed above, we can conclude that if we switch primal and dual cochains, PEC is implied instead of PMC. More specifically, we can set $\bm E$, $\bm B$ as dual cochains and $\bm H$, $\bm D$, $\bm J$ as primal cochains instead. Then equation of cochain $\bm H$ without extra boundary condition will imply PEC boundary condition.

Especially for $3$-D equation (\ref{eqn:e_3d}), although PEC boundary condition can be implemented simply by ignoring the boundary elements, it will result in considerable error. Instead, we can write equivalent equation for cochain $\bm H$ as
\begin{align}
\label{eqn:3dpec}
\left[ (\mat{d}^{(1)})^T \star^{(2)}_{\epsilon^{-1}} \mat{d}^{(1)} \right] \bm H = k_0^2 \left[\star^{(1)}_{\mu} \right]\bm H + \left[(\mat{d}^{(1)})^T \star^{(2)}_{\epsilon^{-1}} \right] \bm J.
\end{align}
Here, $\star^{(2)}_{\epsilon^{-1}}$ and $\star^{(1)}_{\mu}$ are defined in a similar way with $\star^{(2)}_{\mu^{-1}}$ and $\star^{(1)}_{\epsilon}$. Then, without extra boundary condition inserted on (\ref{eqn:3dpec}), PEC is implicitly implemented.

This approach can also applies to $2$-D problems. Instead of solving Equation (\ref{eqn:te}) of dual $0$-cochain $\bm H_z$ with PEC boundary condition for TE modes, we can set $\bm H_z$ as primal $0$-cochain and rewrite Equation (\ref{eqn:31}) with DEC as
\begin{equation}
\label{eqn:65}
\left[(\mat{d}^{(0)})^T \star^{(1)} \mat{d}^{(0)} \right] \bm H_z = k_s^2 \left[ \star^{(0)}\right] \bm H_z.
\end{equation}

In the language of PDE, by switching primal and dual cochains, PEC boundary condition is changed from an essential boundary condition to natural boundary condition.

\subsection{Periodic boundary condition}
\begin{figure}[ht]
\centering
\includegraphics[width = 0.4\textwidth]{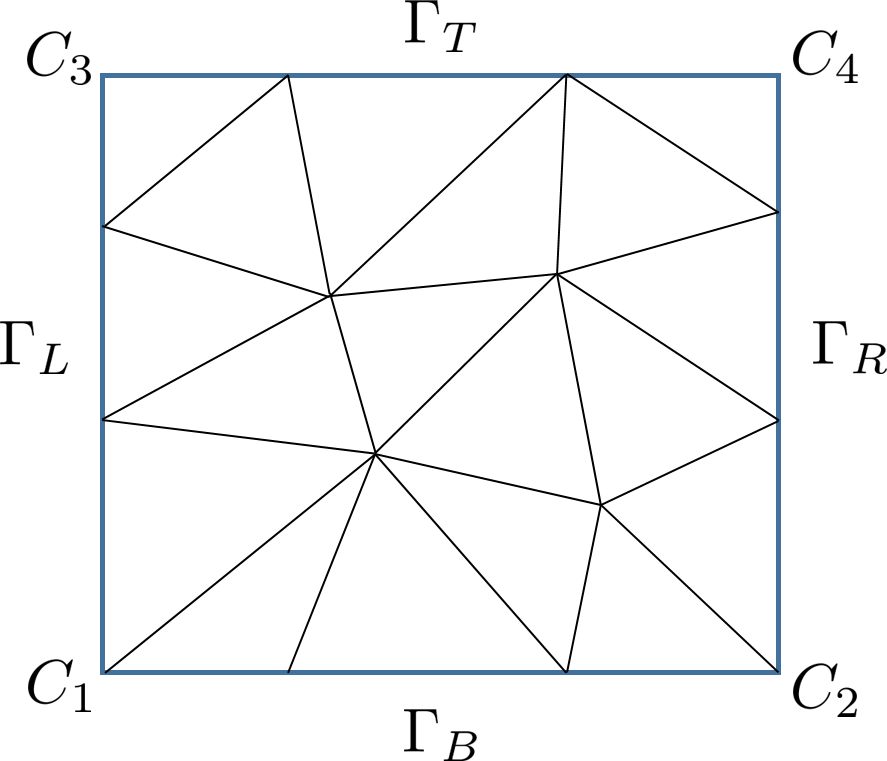}
\caption{A simple periodic triangular mesh for a square unit cell. Here, $C_1, C_2, C_3, C_4$ are four vertices, and $\Gamma_L, \Gamma_B, \Gamma_R, \Gamma_T$ are four boundary edges.}
\label{fig:p_s}
\end{figure}

Periodic boundary condition can also be implemented in our method with a procedure similar to FEM \cite{jin2015}. For a periodic structure with square unit cell, shown in Figure \ref{fig:p_s}.

A projection matrix $\mat{P}$ is needed to reduce unknowns on the boundary. If we use subscript $I$ to denote vertices strictly inside, subscripts $L$, $R$, $B$, $T$ for four boundaries, and subscripts $C_i$ ($i = 1,2,3,4$) for four corners, then the cochain $\bm E_z$ can be divided into nine components
\begin{equation}
\bm E_z = \{ \bm E_{z,I}; \bm E_{z,L}; \bm E_{z,R}; \bm E_{z,B}; \bm E_{z,T}; \bm E_{z,C_1}; \bm E_{z,C_2}; \bm E_{z,C_3}; \bm E_{z,C_4} \}
\end{equation}
This can be reduced to a vector only containing independent unknowns by applying periodic boundary conditions:
\begin{equation}
\bm E_z = \left[\mat{P}\right] \bm E_{z,\text{reduced}},
\end{equation}
where
\begin{equation}
\bm E_{z,\text{reduced}} \triangleq \{ \bm E_{z,I}; \bm E_{z,L}; \bm E_{z,B}; \bm E_{z,C_1} \}
\end{equation}
and
\begin{equation}
\mat{P} \triangleq 
\begin{pmatrix}
I & 0 & 0 & 0 \\
0 & I & 0 & 0 \\
0 & Ie^{-i\Psi_x} & 0 & 0 \\
0 & 0 & I & 0 \\
0 & 0 & Ie^{-i\Psi_y} & 0 \\
0 & 0 & 0 & 1 \\
0 & 0 & 0 & e^{-i\Psi_x} \\
0 & 0 & 0 & e^{-i\Psi_y} \\
0 & 0 & 0 & e^{-i(\Psi_x + \Psi_y)}
\end{pmatrix}
\end{equation}
where $\Psi_x = k_x L$ and $\Psi_y = k_y L$ are the phase shift between adjacent unit cells along $x$-axis and $y$-axis. It should be noted that operator $\mat P$ will have different form for lattice with different unit cell.
Then with periodic boundary condition, Equation (\ref{eqn:tm_in}) for TM modes can be written as
\begin{equation}
\left[\mat P \right]^{\dagger} \left[(\mat{d}^{(0)})^T \star^{(1)}_{\mu^{-1}} \mat{d}^{(0)} \right] \left[\mat P \right] \bm E_{z,\text{reduced}} = k_0^2 \left[ \mat P \right]^{\dagger}  \left[ \star^{(0)}_{\epsilon}\right] \left[ \mat P \right] \bm E_{z,\text{reduced}}.
\label{eqn:tm_per}
\end{equation}

For TE modes, using the same technique introduced in PEC boundary condition, we can place $0$-cochain $\bm H_z$ on primal mesh, and defined $\bm H_{z, \text{reduced}}$ in the same way as above. Then equation for TE modes with periodic boundary condition is written as
\begin{equation}
\left[\mat P \right]^{\dagger} \left[(\mat{d}^{(0)})^T \star^{(1)}_{\epsilon^{-1}} \mat{d}^{(0)} \right] \left[\mat P \right] \bm H_{z,\text{reduced}} = k_0^2 \left[ \mat P \right]^{\dagger}  \left[ \star^{(0)}_{\mu}\right] \left[ \mat P \right] \bm H_{z,\text{reduced}}.
\label{eqn:te_per}
\end{equation}
Therefore, TM and TE band diagrams in photonic crystals can be obtained by solving Equations (\ref{eqn:tm_per}) and (\ref{eqn:te_per}) with different $k_x$ and $k_y$.

\subsection{Two-dimensional ABCs}
Absorbing boundary conditions (ABCs) can also be implemented easily, and they represent inhomogeneous natural (Neumann) boundary conditions. For first-order ABC, we should have a proportional relation between tangential component of $\v H_s(\v r)$ and $E_z(\v r)\hat{z}$ \cite{chew1994, jin2015}.
\begin{equation}
\hat{n} \times \v H_s \approx \frac{1}{i\omega\mu} \left(i k_s + \frac{\kappa}{2} \right) E_z \hat{z} = \left( \frac{1}{\eta} + \frac{\kappa}{i 2 \omega\mu} \right) E_z \hat{z}
\end{equation}
Here $\eta = \sqrt{\frac{\epsilon}{\mu}}$ is the impedance, and $\kappa$ is the curvature at the boundary ($0$ for flat boundary). Then the last two terms of Equation (\ref{eqn:49}) can be approximated as
\begin{equation}
H_{s, \overline{e_1 p_2}} + H_{s, \overline{p_2 e_2}} = |\c L_{p_2}^{\partial}| \left( \frac{1}{\eta} + \frac{\kappa_{p_2}}{i 2 \omega\mu} \right) E_{z,p_2}.
\end{equation}
Here, $|\c L_{p_2}^{\partial}| = |\overline{e_1 p_2}| + |\overline{p_2 e_2}|$ is the total length of dashed lines in Figure \ref{fig:bc2}, and superscript $\partial$ stands for boundary. Then with first-order ABC, (\ref{eqn:tm}) should be adjusted as:
\begin{equation}
\label{eqn:2abc1st}
\left[ (\mat{d}^{(0)})^T \star^{(1)} \mat{d}^{(0)} \right] \bm E_z + \left[i k_s + \frac{1}{2} \star^{\kappa} \right] \cdot \left[ \star^{\partial} \right] \bm E_z = k_s^2 \left[ \star^{(0)} \right] \bm E_z,
\end{equation}
where
\begin{equation}
\star^{\kappa} =   
\begin{bmatrix}
\ddots & & & \\
& |\c \kappa_i| & & \\
& & \ddots & \\
& & & \scalebox{2}{$\mat{0}$}
\end{bmatrix}, \
\star^{\partial} =  
\begin{bmatrix}
\ddots & & & \\
& |\c L_i^{\partial}| & & \\
& & \ddots & \\
& & & \scalebox{2}{$\mat{0}$}
\end{bmatrix}.
\end{equation}
Here, $\star^{\kappa}$ and $\star^{\partial}$ are constructed as $N_0 \times N_0$ matrices with nonzero diagonal elements only for boundary vertices, and subscript $i$ refers to a vertex on the boundary. Therefore, they are highly sparse matrices and only act on boundary elements.

Second-order ABC is derived in \cite{chew1995,jin2015} as
\begin{equation}
-i\omega \mu (\hat{n} \times \v H_s) \approx \left(-i k - \frac{\kappa}{2} + \frac{\kappa^2}{8 i k} + \frac{\kappa^3}{8k^2} \right) E_z \hat{z} + \left(\frac{1}{2i k} + \frac{\kappa}{2k^2} \right) \frac{\partial^2 E_z}{\partial s^2} \hat{z}
\label{eqn:57}
\end{equation}
Here $\frac{\partial^2 E_z}{\partial s^2}$ refers to second-order derivative on the boundary. The first term of (\ref{eqn:57}) can be implemented the same way as in first-order ABC. To interpret the second term of (\ref{eqn:57}) with DEC, we introduce the derivative operator $\mat{d}^{(0, \partial)}$ and Hodge star operator $\star^{1, \partial}$ confined on the boundary. Here $\mat{d}^{(0, \partial)}$ is only nonzero for boundary primal edge and point elements, and is defined the same way as $\mat{d}^{(0)}$, while $\star^{1, \partial}$ is defined as
\begin{equation}
\star^{1, \partial} \triangleq  
\begin{bmatrix}
\ddots & & & \\
& \frac{1}{|l_j|} & & \\
& & \ddots & \\
& & & \scalebox{2}{$\mat{0}$}
\end{bmatrix}.
\end{equation}
Here subscript $l_j$ refers to a primal edge on the boundary. Then $\frac{\partial^2 E_z}{\partial s^2}$ corresponds to a dual $1$-cochain only on the boundary. More specifically, as shown in Figure \ref{fig:bc2}, we can derive
\begin{equation}
\begin{split}
\int_{e_1 \rightarrow p_2 \rightarrow e_2} \frac{\partial^2 E_z}{\partial s^2} dl & = \frac{\partial E_z}{\partial s}\bigg|_{e_2} - \frac{\partial E_z}{\partial s}\bigg|_{e_1} \\
& \approx \frac{E_{z,p_3} - E_{z,p_2}}{|p_2 p_3|} - \frac{E_{z,p_2} - E_{z,p_1}}{|p_1 p_2|}.
\end{split}
\end{equation}
Therefore, $\frac{\partial^2 E_z}{\partial s^2}$ can be represented with dual $1$-cochain $[(\mat{d}^{(0, \partial)})^T \star^{1,\partial} \mat{d}^{(0, \partial)} ] \bm E_z$. Then with second-order ABC, (\ref{eqn:tm}) is adjusted as
\begin{equation}
\label{eqn:2abc2nd}
\begin{split}
&\left[ (\mat{d}^{(0)})^T \star^{(1)} \mat{d}^{(0)} \right] \bm E_z + \left[i k_s + \frac{1}{2} \star^{\kappa} - \frac{(\star^{\kappa})^2}{8 i k} - \frac{(\star^{\kappa})^3}{8 k^2} \right] \cdot \left[ \star^{\partial} \right] \bm E_z \\
&+ \left[ \frac{1}{2 i k} + \frac{\star^{\kappa}}{2 k^2} \right] \left[ (\mat{d}^{(0, \partial)})^T \star^{(1,\partial)} \mat{d}^{(0, \partial)} \right] \bm E_z  = k_s^2 \left[ \star^{(0)} \right] \bm E_z.
\end{split}
\end{equation}

\subsection{Three-dimensional ABCs}
\begin{figure}[htbp]
\centering
\includegraphics[width = 0.9\textwidth]{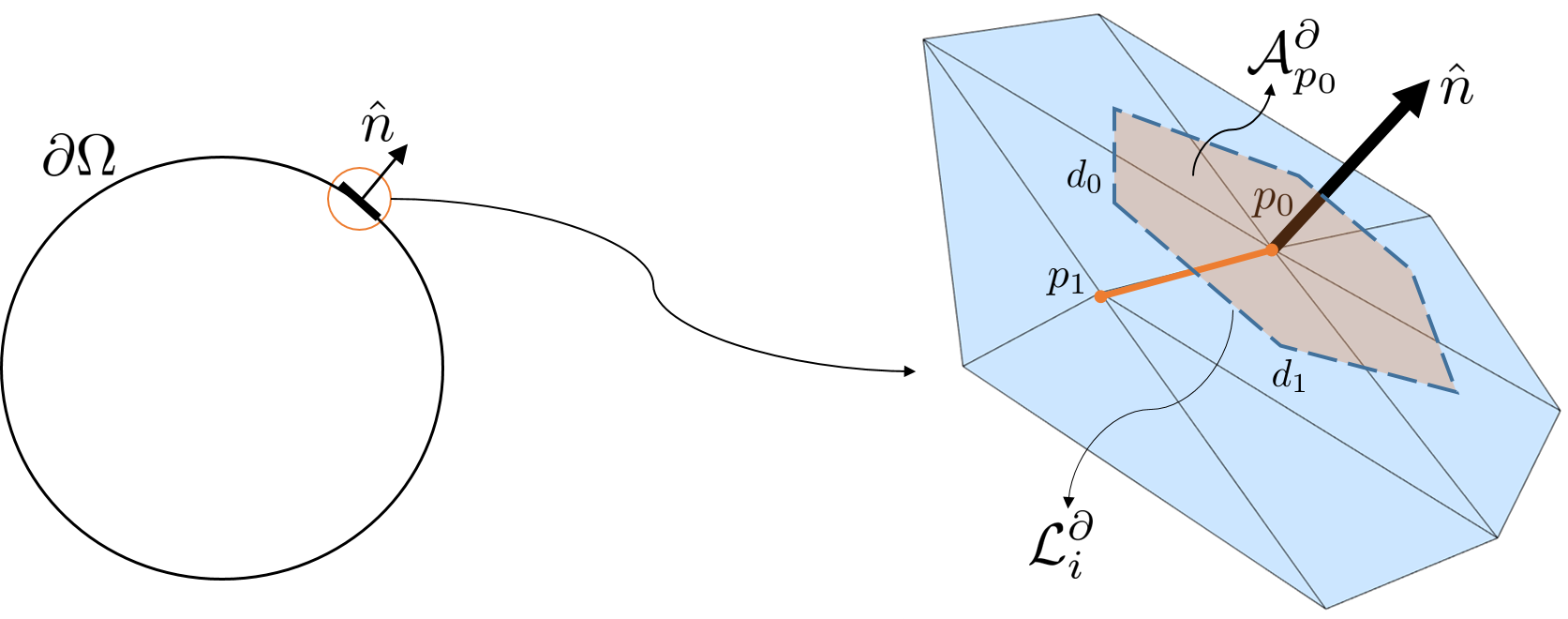}
\caption{Left is a computational domain with surface $\partial \Omega$. Right is the simplicial mesh at the circled region of the right figure. Here primal edge $\overline{p_0 p_1}$ is denoted by $l_i$, $\overline{d_0 d_1}$ is the surface dual edge $\c L_{i}^{\partial}$, and the area on surface enclosed by dashed lines is the dual face $\c A_{p_0}^{\partial}$ associated with $p_0$.}
\label{fig:bc3}
\end{figure}

For a $3$-D case, the simplest ABC is the Sommerfeld radiation condition \cite{jin2015} 
\begin{equation}
\hat{n} \times (\nabla \times \v E) = i\omega \mu_0 (\hat{n} \times \v H) \approx -ik_0 \hat{n}\times (\hat{n}\times \v E)
\end{equation}
This relation can be implemented in a similar way as in above $2$-D case. With this first-order ABC, (\ref{eqn:e_3d}) is adjusted as
\begin{align}
\label{eqn:3d1stabc}
\left[ (\mat{d}^{(1)})^T \star^{(2)}_{\mu^{-1}} \mat{d}^{(1)} \right] \bm E + i k_0 \left[ \star^{(1, \partial)} \right] \bm E = k_0^2 \left[\star^{(1)}_{\epsilon} \right]\bm E + i\omega \bm J,
\end{align}
with
\begin{equation}
\star^{(1, \partial)} \triangleq 
\begin{bmatrix}
\ddots & & & \\
& \frac{|\c L_i^{\partial}|}{|l_i|} & & \\
& & \ddots & \\
& & & \scalebox{2}{$\mat{0}$}
\end{bmatrix}.
\end{equation}
Similarly, $\star^{(1, \partial)}$ is a $N_1 \times N_1$ matrix only containing diagonal elements for boundary edges, and $\c L_i^{\partial}$ is the surface dual edge shown in Figure \ref{fig:bc3}.

The second-order ABC \cite{webb1989} requires the electric field on the boundary to satisfy 
\begin{equation}
\label{eqn:76}
\hat{n} \times (\nabla \times \v E) \approx - i k_0 \hat{n} \times \hat{n} \times \v E + \beta \nabla \times \left[ \hat{n} \hat{n} \cdot (\nabla \times \v E) \right] + \beta \nabla_t (\nabla_t \cdot \v E)
\end{equation}
where $\nabla_t$ is the surface tangential gradient operator, the subscript $n$ represents the normal component on the surface, and parameter $\beta$ is defined as
\begin{equation*}
\beta \triangleq \frac{1}{2( i k_0 + \kappa)}
\end{equation*}
where $\kappa$ is the curvature of the boundary surface (zero for flat surfaces).

While in the second term of (\ref{eqn:76}), $\hat{n}\hat{n} \cdot (\nabla \times \v E)$, or $\hat{n} (\nabla \times \v E)_n$ can be simply represented by $\mat{d}^{(1, \partial)} \bm E$, where $\mat{d}^{(1,\partial)}$ is defined the same way with $\mat{d}^{(1)}$ only containing relation between boundary surface faces and edges. It should be noted that on the $2$-D surface $\partial \Omega$,  $\hat{n} (\nabla \times \v E)_n$ leads to a primal $2$-cochain $\mat{d}^{(1,\partial)} \bm E$. Therefore, for the second curl operator $\nabla \times$ to operate on this primal $2$-cochain, a Hodge star operator $\star^{2, \partial}$ needs to be implemented first. Here $\star^{(2, \partial)}$ is defined as
\begin{equation}
\star^{(2, \partial)} \triangleq 
\begin{bmatrix}
\ddots & & & \\
& \frac{1}{|A_i^{\partial}|} & & \\
& & \ddots & \\
& & & \scalebox{2}{$\mat{0}$}
\end{bmatrix}
\end{equation}
where surface primal face $A_i^{\partial}$ represents a triangle on boundary surface. Then $\nabla \times$ corresponds to $(\mat{d}^{(1, \partial)})^T \star^{(2, \partial)}$, because
\begin{equation}
\int_{d_0}^{d_1} \left[ \nabla \times \hat{n} (\nabla \times \v E)_n \right] \cdot \hat{n}\times d\v l = (\nabla \times \v E)_n \bigg|_{d_1} - (\nabla \times \v E)_n \bigg|_{d_0}.
\end{equation}

In the third term of (\ref{eqn:76}), with the language of DEC, surface divergence $\nabla_t \cdot$ is denoted by the transpose of $\mat{d}^{(0, \partial)}$ and a surface Hodge star operator $\star^{(1, \partial)}$ together acting on primal $1$-cochain $\bm E$. Then $[ (\mat{d}^{(0, \partial)})^T \star^{(1, \partial)} ] \bm E$ represents a dual $2$-cochain on boundary surface. For the surface gradient $\nabla_t$ to operate appropriately, another Hodge star operator $(\star^{(0, \partial)})^{-1}$ needs to be inserted to map this dual $2$-cochain to primal $0$-cochain. Hodge star $\star^{(0, \partial)}$ is defined only for surface points as
\begin{equation}
\star^{(0, \partial)} \triangleq
\begin{bmatrix}
\ddots & & & \\
& |\c A_{p_0}^{\partial}| & & \\
& & \ddots & \\
& & & \scalebox{2}{$\mat{0}$}
\end{bmatrix}
\end{equation}
where $\c A_{p_0}^{\partial}$ is illustrated in Figure \ref{fig:bc3}.
Then surface gradient $\nabla_t$ can be represented by $\mat{d}^{(0, \partial)}$, and $[ \mat{d}^{(0, \partial)} (\star^{(0, \partial)})^{-1} (\mat{d}^{(0, \partial)})^T \star^{(1, \partial)} ] \bm E$ is a primal $1$-cochain. However, since $\hat{n} \times (\nabla \times \v E)$ (tangential component of $\v H$ field), the left hand side of Equation (\ref{eqn:76}), normally is denoted by a dual $1$-cochain, Hodge star $\star^{(1, \partial)}$ needs to be implemented again in this third term.

Therefore, Equation (\ref{eqn:e_3d}) with second-order ABC is written as
\begin{equation}
\begin{split}
& \left[ (\mat{d}^{(1)})^T \star^{(2)}_{\mu^{-1}} \mat{d}^{(1)} \right] \bm E + i k_0 \left[ \star^{(1,\partial)} \right] \bm E + \beta \left[ (\mat{d}^{(1, \partial)})^T \star^{(2, \partial)} \mat{d}^{(1, \partial)} \right] \bm E \\
& + \beta \left[ \star^{(1, \partial)} \mat{d}^{(0, \partial)} (\star^{(0, \partial)})^{-1} (\mat{d}^{(0, \partial)})^T \star^{(1, \partial)} \right] \bm E = k_0^2 \left[\star^{(1)}_{\epsilon} \right]\bm E + i\omega \bm J
\end{split}
\end{equation}
where we assume that the curvature is a constant on boundary surface for simplicity.

\section{Numerical Examples}
\label{sc5}
\subsection{Validation: homogeneous circular waveguide}

\begin{figure}[H]
\centering
\includegraphics[width = 0.6\textwidth]{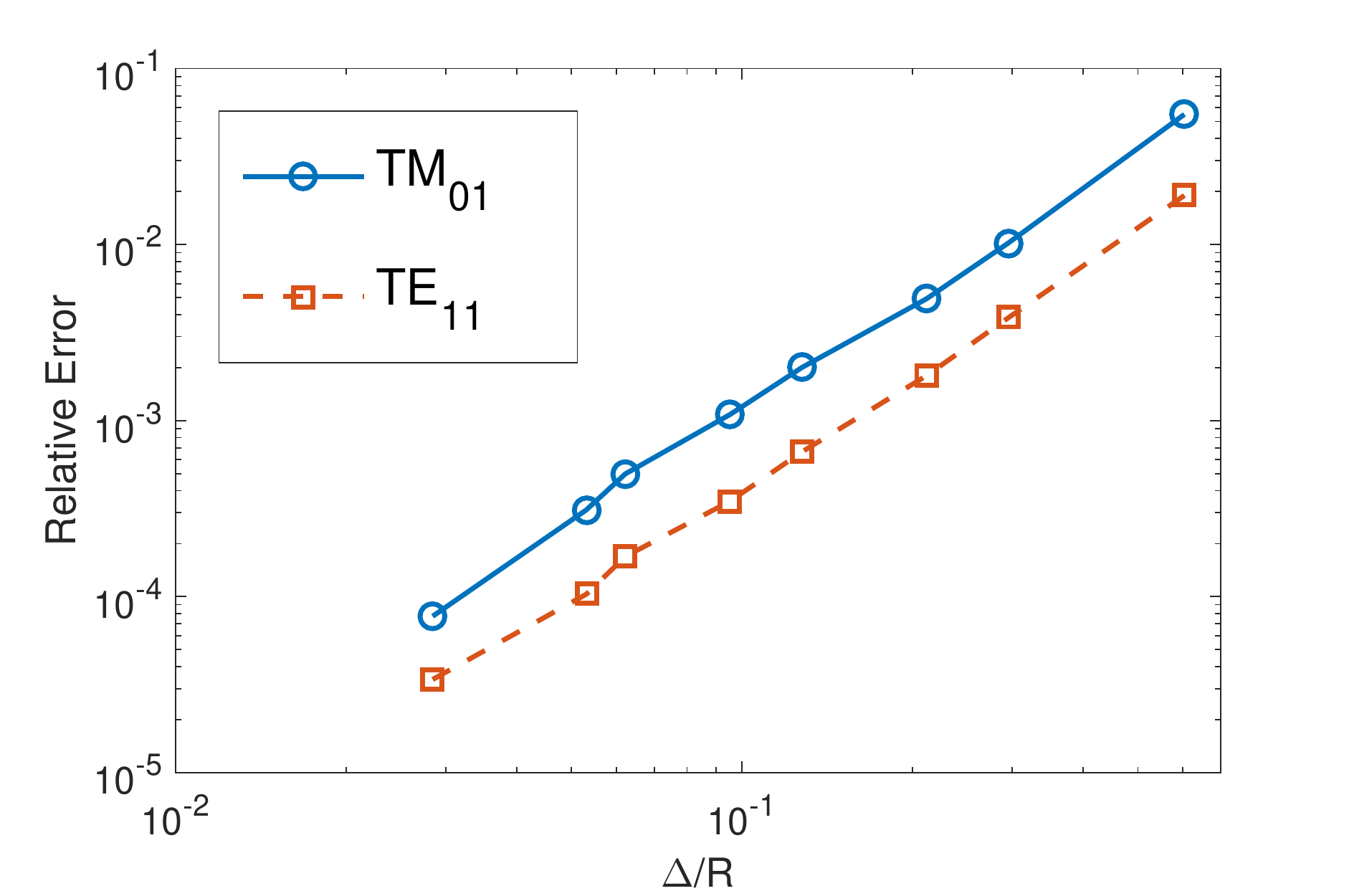}
\caption{Relative error vs.\ $\Delta/R$ for a hollow circular waveguide, where $R$ is the radius of the waveguide.}
\label{fig:TM_cwg_cvg}
\end{figure}

\begin{figure}[H]
\centering
\includegraphics[width = \textwidth]{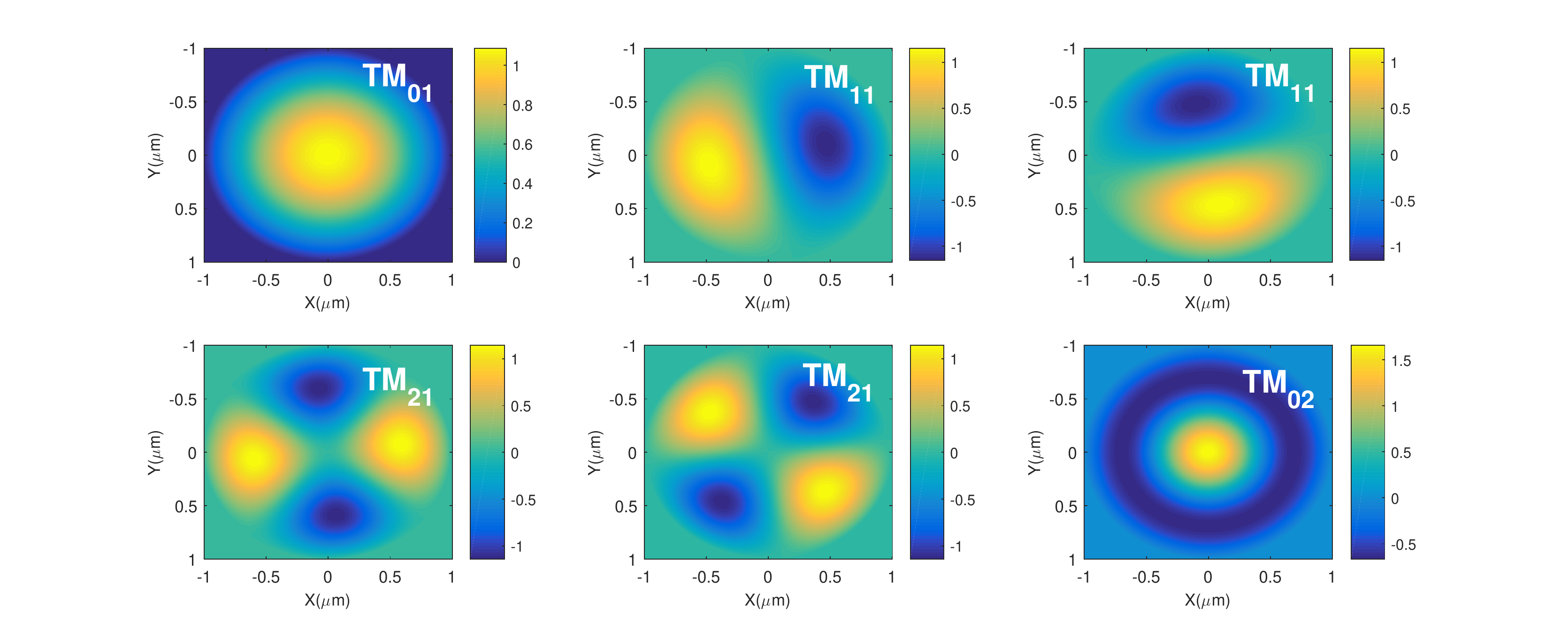}
\caption{First 6 TM modes in a circular waveguide with PEC boundary condition.}
\label{fig:TM_cwg}
\end{figure}

Then we can solve for the TM and TE modes in a hollow circular waveguide by using (\ref{eqn:63}) and (\ref{eqn:65}). The analytical value of $k_s$ for the TM and TE modes are roots of Bessel functions and roots of derivatives of Bessel functions. More specifically, $k_s = 2.40482555769577$ for TM$_{01}$ mode, and $k_s = 1.84118378134065$ for TE$_{11}$ mode. Comparing mesh with different fineness, the relation between relative error and maximum edge length $\Delta$ can be plotted as in Figure \ref{fig:TM_cwg_cvg}. A fitting shows that the convergence order is $2.0932$ for TM$_{01}$ mode and $2.0689$ for TE$_{11}$ mode. \footnote{This second order convergence will be proved in our future publication.}
The profiles of $E_z (\v r)$ field for first six TM modes in a circular waveguide are plotted in Figure \ref{fig:TM_cwg}.

\subsection{Microstructured optical fibers}
\begin{figure}[htbp]
	\begin{subfigure}{.5\textwidth}
		\centering
		\includegraphics[width = 0.6\textwidth]{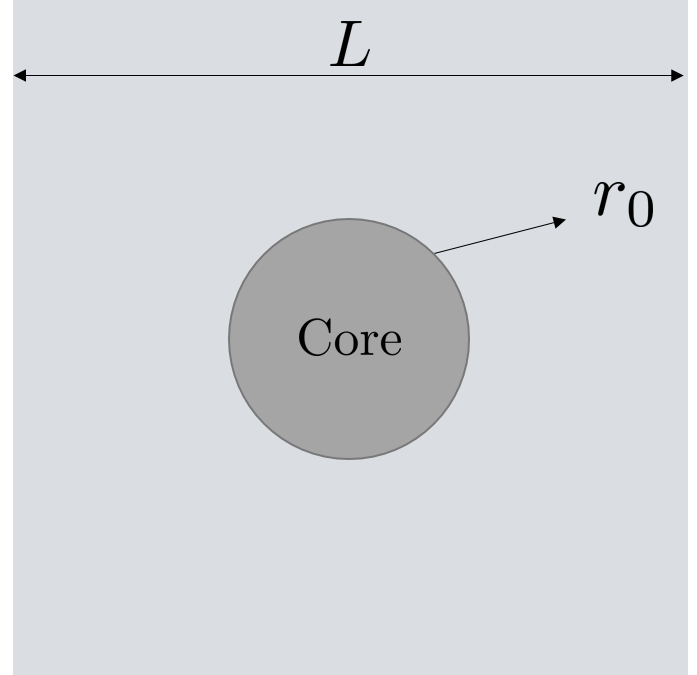}
		\caption{}
		\label{fig:sif}
	\end{subfigure}
	\begin{subfigure}{.5\textwidth}
		\centering
		\includegraphics[width = 0.6\textwidth]{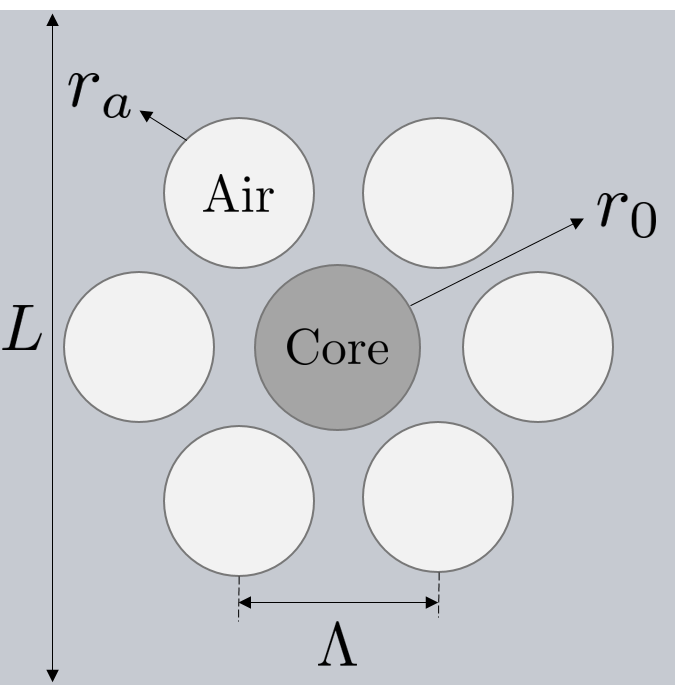}
		\caption{}
		\label{fig:ahaof}
	\end{subfigure}
\caption{(a) Step-index optical fiber with radius $r_0 = 3\mu$m and refractive index $1.45$ at wavelength$= 1.5\mu$m, and $L=12\mu$m; (b) Structure of Air-hole assisted optical fiber with parameters: core index $1.45$, silica cladding index $1.42$, $r_0 = 2\mu$m, $r_a= 2\mu$m, $\Lambda = 5\mu$m, and $L = 16\mu$m.}
\label{fig:msf}
\end{figure}

Using Equation (\ref{eqn:38_pec}), we investigate the fundamental effective index of a step-index optical fiber and a air-hole assisted optical fiber (AHAOF), shown in Figure \ref{fig:msf}. The results are compared to both analytical solution (step-index fiber) and numerical solution by finite difference method \cite{zhu2002}. 
 
\begin{figure}[H]
\centering
\includegraphics[width=0.6\textwidth]{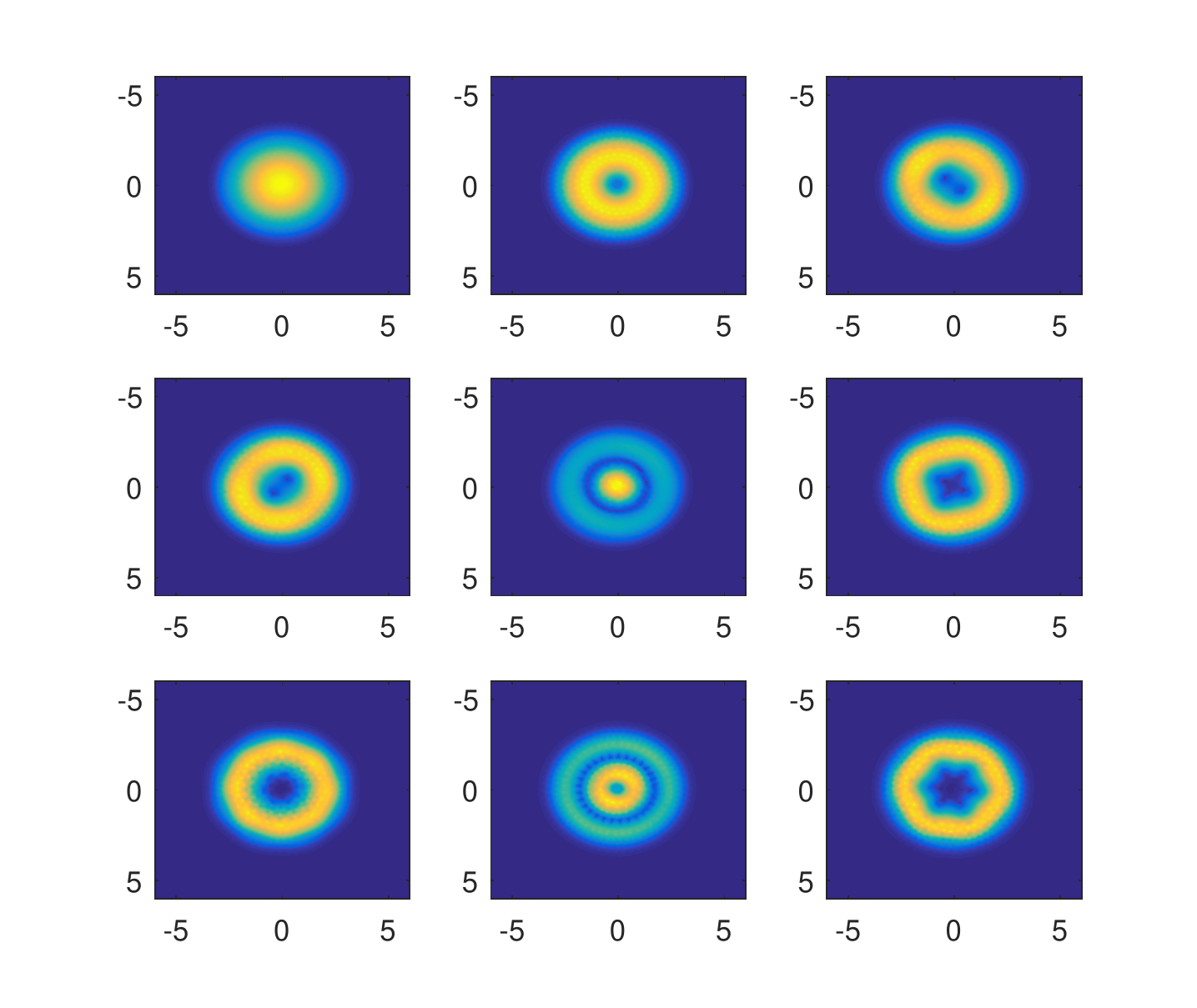}
\caption{Intensity of transverse electrical field of first $9$ modes.}
\label{fig:optf}
\end{figure}

The step-index optical fiber has parameters shown in Figure \ref{fig:sif}. It is surrounded with air (refractive index $=1$). The fundamental mode index is defined as
\begin{equation*}
n_{\text{eff}} = \frac{k_z}{k_0} = \frac{\beta}{k_0}
\end{equation*}
Then, the fundamental mode index can be calculated analytically as $n_{\text{eff}}=1.438604$. This step-index optical fiber can also be viewed as an inhomogeneous waveguide, and its modes can be solved by Equation (\ref{eqn:38_pec}) with PEC boundary condition. Our numerical solution gives the fundamental mode index $n_{\text{eff}} = 1.4386043519$ with 2,972 triangles and 1,561 vertices, which agrees very well with the analytical value. The transverse electric field intensity of first $9$ non-degenerate modes are plotted in Figure \ref{fig:optf}. There are two main methods to map a $1$-cochain to a vector field. One is to assume the vector field is constant in each triangle patch \cite{hirani2003}; the other is to expand the vector field with Whitney forms \cite{desbrun2005, desbrun2008, moon2015}.

\begin{figure}[H]
\centering
\includegraphics[width=0.8\textwidth]{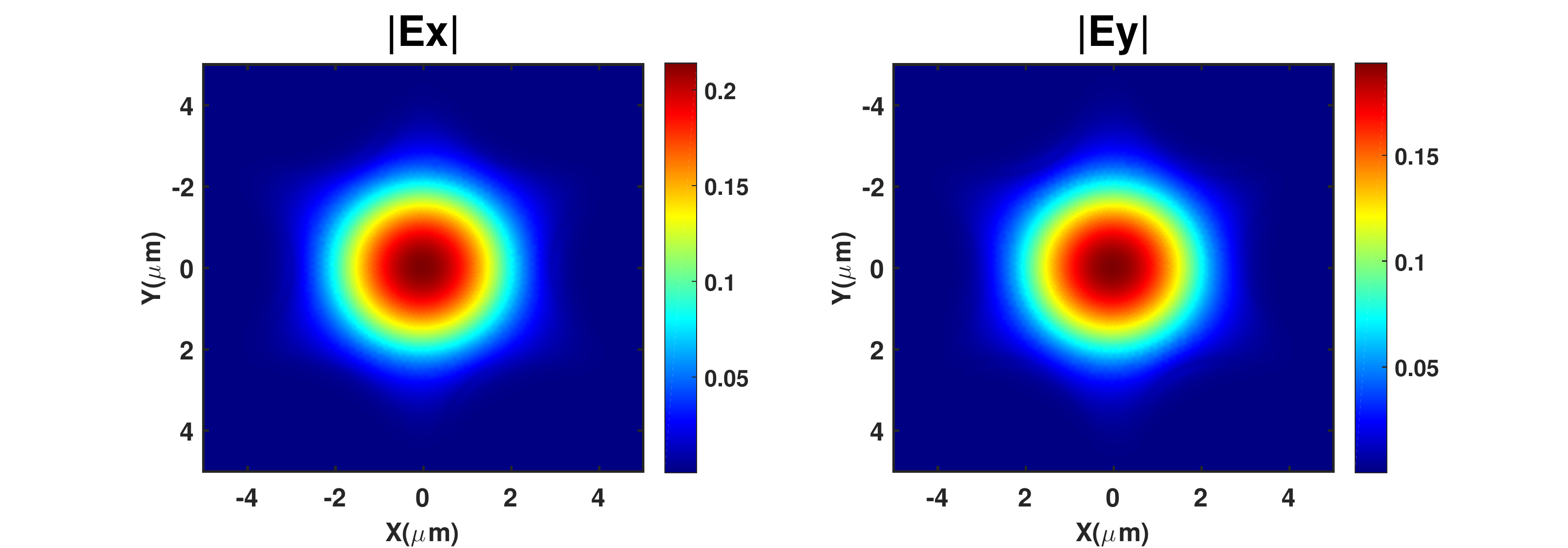}
\caption{$|E_x|$ and $|E_y|$ profile of AHAOF's fundamental mode.}
\label{fig:ahaof_1st}
\end{figure}

The structure of AHAOF we have considered is shown in Figure \ref{fig:ahaof}. The guiding core is surrounded by air-holes. The advantage of this structure compared to normal optical fibers is that its dispersion is easily tailorable. Literature \cite{zhu2002} used a $120\times 120$ grid for a quadrant window with 28,800 unknowns and solved the fundamental mode index as $n_{\text{eff}} = 1.4353602$. We adopted a triangular mesh for the entire squire domain with 4,240 triangles and 6,400 edges (number of unknowns), and obtained $n_{\text{eff}} = 1.4353696$. The intensity of $|E_x|$ and $|E_y|$ are plotted in Figure \ref{fig:ahaof_1st}, and the profiles reflect the position of surrounding air-holes.

\subsection{Photonic crystals}
Photonic crystal, a periodic optical nanostructure, has very broad applications. Among these, two dimensional photonic crystals is not only used to produce commercial photonic-crystal fibers, they are also applied to form nano-cavities in quantum optics \cite{yoshie2004}.
\begin{figure}[htbp]
	\begin{subfigure}{0.5\textwidth}
	\centering
	\includegraphics[width = \textwidth]{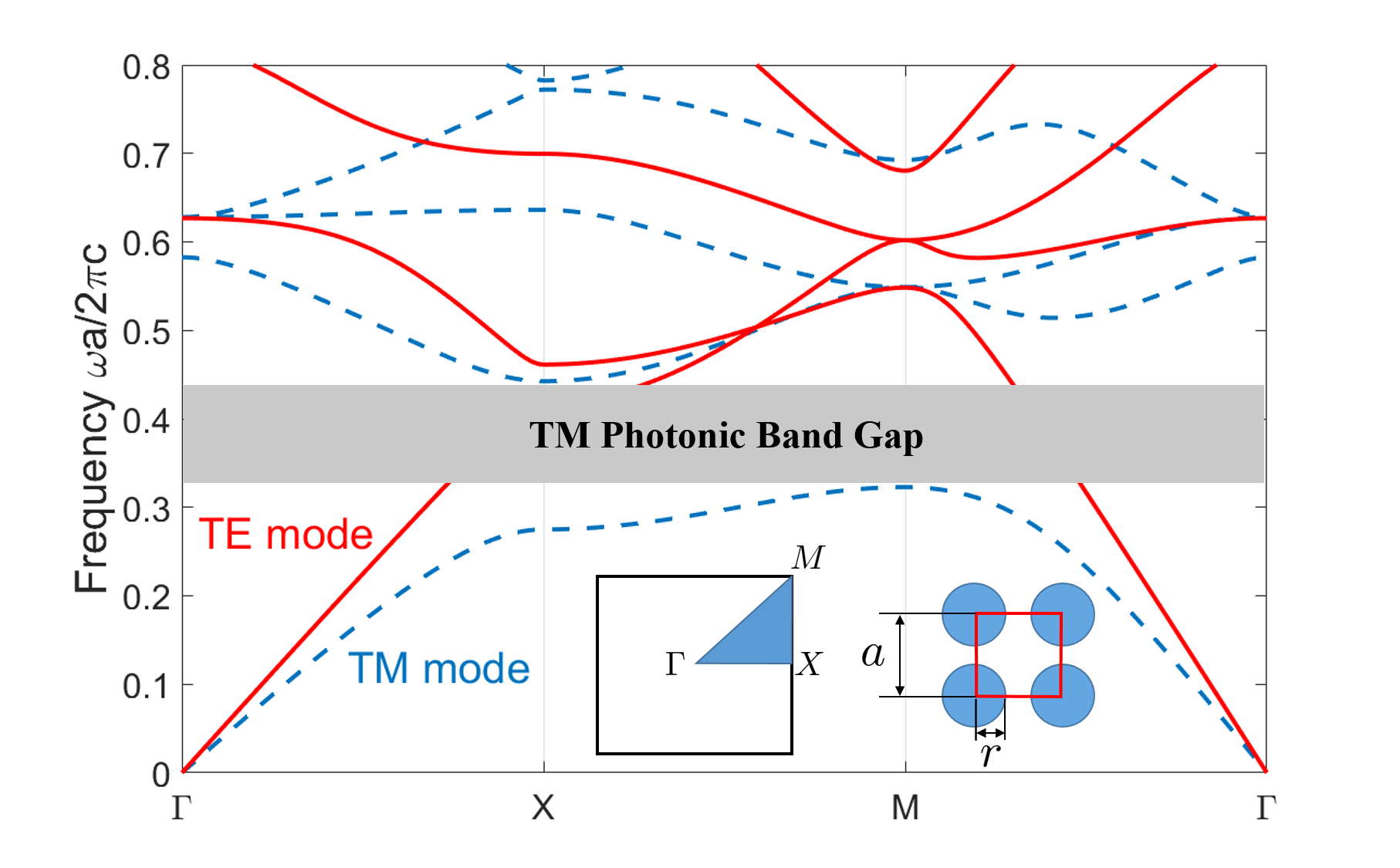}
	\caption{}
	\label{fig:phc_d}
\end{subfigure}
\begin{subfigure}{0.5\textwidth}
	\centering
	\includegraphics[width = \textwidth]{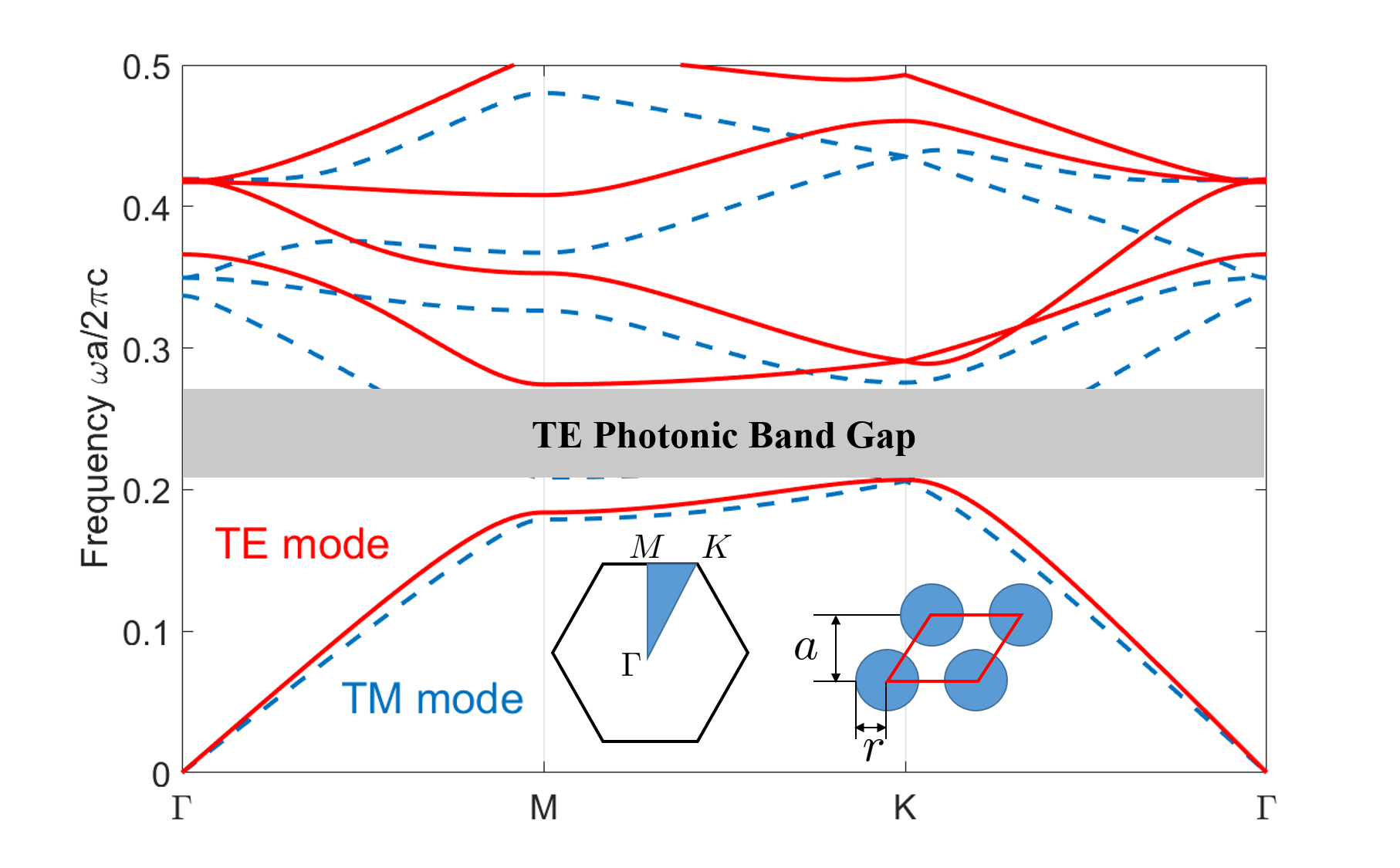}
	\caption{}
	\label{fig:phc_air}
\end{subfigure}
	\caption{(a) The band structures of square lattice photonic crystal. Lattice constant is $a$, the radius
	of cylinders $r = 0.2a$, and the dielectric constant of cylinders $8.9$; (b) The band structure of
	this triangular lattice photonic crystal. Lattice constant is $a$, the radius of air-holes $r = 0.3a$, and the dielectric constant of slab is $13$.}
	\label{fig:phc}
\end{figure}

The band structure of photonic crystals can also be investigated in our framework with periodic boundary condition. By solving an eigen problem in Equations (\ref{eqn:tm_per}) and (\ref{eqn:te_per}) with different $k_x$ and $k_y$, we can obtain a band diagram of a periodic structure. We considered two structures, shown in Figure \ref{fig:phc}. The structure in Figure \ref{fig:phc_d} is dielectric cylinders positioned in squared lattice, and the structure in Figure \ref{fig:phc_air} is a dielectric slab with air-holes placed on a triangular lattice. The band structure shown in Figure \ref{fig:phc} agrees really well with results in reference \cite{jin2015} which uses FEM. Observation shows that the left structure admits a TM photonic band gap, while the right structure has a TE photonic band gap.

\subsection{2-D scattering problems}
\begin{figure}[H]
\centering
\includegraphics[width=0.8\textwidth]{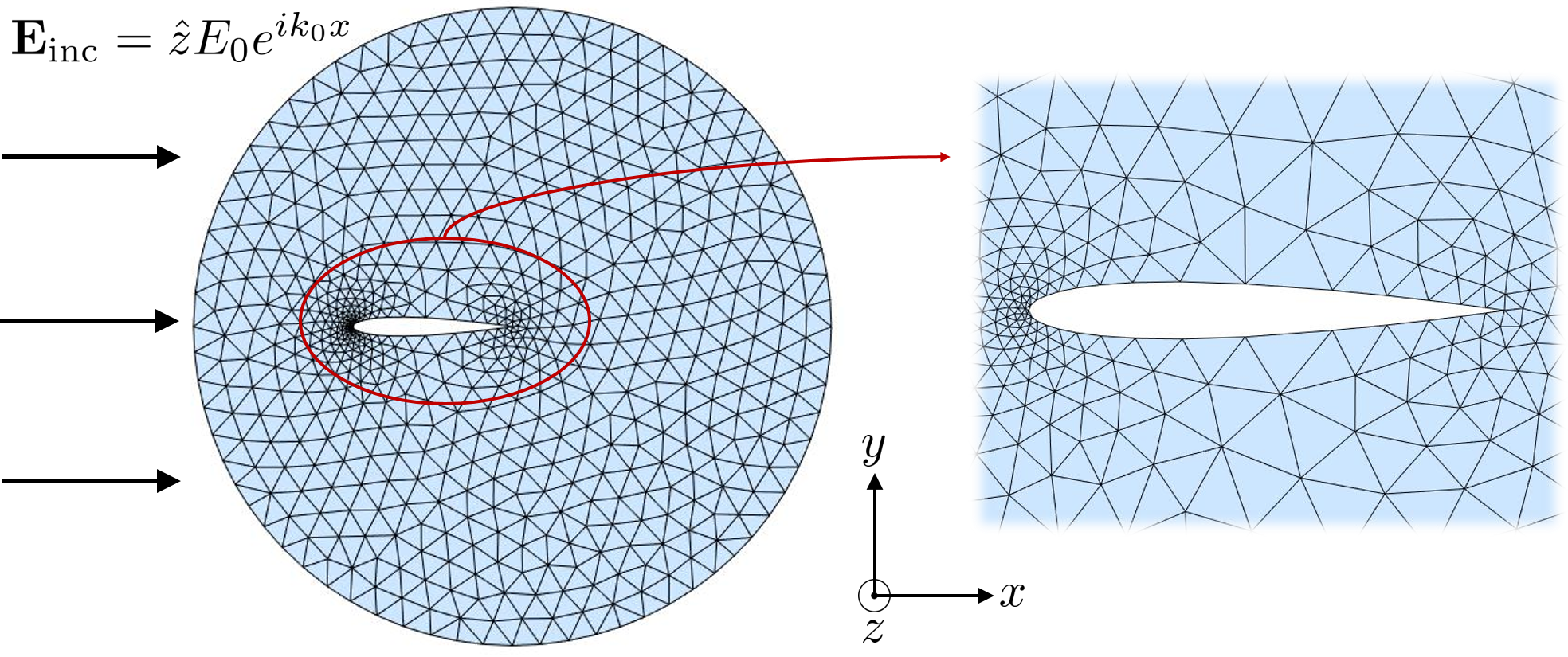}
\caption{An airfoil is scattering by an incident plane wave. The outer boundary is chosen to be centered at the trailing edge with radius twice the length of this airfoil.}
\label{fig:af}
\end{figure}

\begin{figure}[H]
\centering
\includegraphics[width=\textwidth]{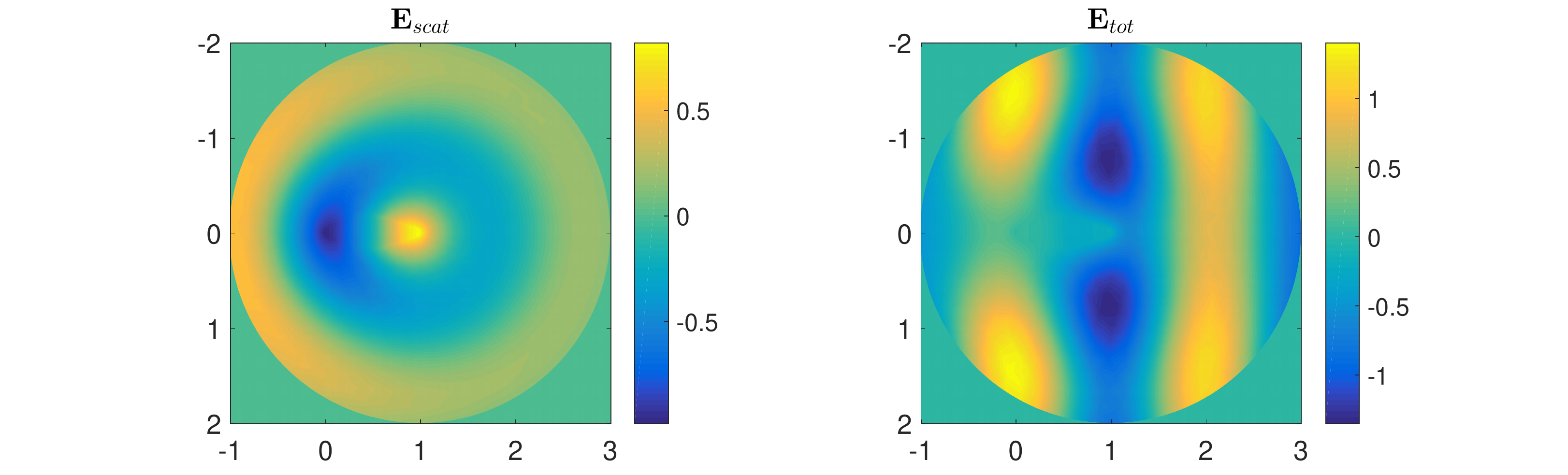}
\caption{Profile of scattered field and total field. The incident wavenumber is $k_0 = \frac{2\pi}{L_0}$, and $L_0$ is the length of the airfoil.}
\label{fig:af_scat}
\end{figure}

Here we investigate the problem of open region scattering by a $2$-D perfect electrical conductor. We have already formulated the Dirichlet boundary condition at the conductor surface with an known incident field $\v E_{\text{inc}} = \hat{z} E_0 e^{i k_0 x}$ as in Equation (\ref{eqn:tm_scat}). We can adopt first-order or second-order ABC, as in Equations (\ref{eqn:2abc1st}) and (\ref{eqn:2abc2nd}), for the outer truncation boundary. The structure we consider is NACA0012 airfoil, shown in Figure \ref{fig:af}.
Then the scattered field and total field with second-order ABC can be plotted as in Figure \ref{fig:af_scat}.

\subsection{Resonant cavities}
Modal analysis can be used to determine the natural resonant frequencies and mode shapes of a structure in a broad field, such as structural mechanics, acoustics and electromagnetics \cite{dai2014}.
\begin{figure}[H]
\centering
\includegraphics[width=0.8\textwidth]{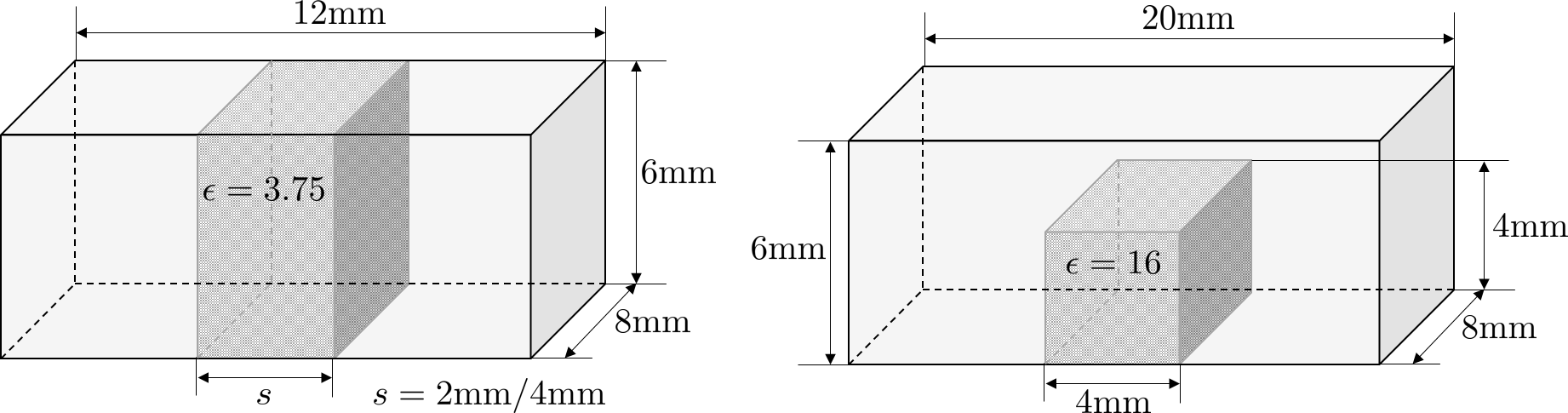}
\caption{Left is a LES-mode resonator and right is a hybrid-mode resonator.}
\label{fig:lse}
\end{figure}

By solving the eigen problem shown in Equation (\ref{eqn:3dpec}) with $\bm J = 0$, we can obtain resonant frequencies of a resonator with conducting enclosures. We first examine an LES-mode resonator and a hybrid-mode resonator with structures introduced in \cite{perepelitsa1997} as shown in Figure \ref{fig:lse}. They are both inhomogeneous cavities. The comparison between current work and results from \cite{perepelitsa1997} used FEM is presented in Table 1.

With ABC boundary condition introduced, we can also calculate the resonant frequency and quality factor of a open structure. However, we need to reformulate the Equation (\ref{eqn:3d1stabc}) as
\begin{equation}
\begin{bmatrix} (\mat{d}^{(1)})^T \star^{(2)}_{\mu^{-1}} \mat{d}^{(1)} & \\ &  \star^{(1)}_{\epsilon} \end{bmatrix} \cdot \begin{pmatrix} \bm E \\ k_0 \bm E \end{pmatrix} 
= k_0 \begin{bmatrix} -i\star^{(1, \partial)} & \star^{(1)}_{\epsilon} \\ \star^{(1)}_{\epsilon} & 0 \end{bmatrix} \cdot \begin{pmatrix} \bm E \\ k_0 \bm E \end{pmatrix} 
\label{eqn:90}
\end{equation}
We applied Equation (\ref{eqn:90}) to find the resonant frequency and the $Q$ factor of the lowest TE mode for a dielectric sphere with radius $r = 160$ $\mu$m and $\epsilon_r = 36$. We placed this sphere in a $320\times320\times320$ $\mu\text{m}^3$ cube and discretized with $23,669$ tetrahedrons. The result is also summarized in Table 1.

\begin{table}[H]
\begin{center}
\caption{Fundamental resonant frequencies (GHz) of inhomogeneous cavities.} \label{tb1}
\begin{tabular}{c||c||c||c||c}
 \hline
 Model & This work  &  \cite{perepelitsa1997} &  \cite{dai2014} & Error (\%) \\
 \hline
 \hline
 LES-mode resonator ($s=2$ mm) & 15.628 & 15.65 & $-$ & 0.14\\
 LES-mode resonator ($s=4$ mm) & 13.346 & 13.35 & 13.34 & 0.03\\
 Hybrid mode resonator & 8.431 & 8.43 & 8,42 & 0.01 \\
 Dielectric sphere* & 152.48 & $-$ & 153.3 & 0.47\\
 \hline
\end{tabular}
\end{center}
\raggedright * The $Q$ factor calculated here for this dielectric sphere is $42.79$, and this value agrees well with $42.31$ in \cite{dai2014}.
\end{table}

\section{Discussion and Conclusions}
In this work, we have adopted discrete exterior calculus (DEC) to formulate and numerically solve various electromagnetic problems in frequency domain. In other words, we have provided an alternative method for computational electromagnetics analysis based on an arbitrary simplicial mesh.

Due to the nature of electromagnetics, the unknown fields are separated into primal cochains and dual cochains. But in practice, we always prefer to solve for the primal cochains, because the error introduced by the boundary can be minimized and the results can be interpolated with Whitney forms. And this is the reason why we treat $H_z(\v r)$ and $\v H(\v r)$ field as primal cochains to solve for TE modes and closed $3$-D problems.

Since DEC keeps the structure and terseness of differential form description of Maxwell's equations, charge continuity relation is exactly preserved, which leads to a great potential in problems involving motions of charged particles \cite{kraus2016}. Another important feature is that all operators acting on cochains are naturally symmetric due to the diagonal Hodge stars. 

In fact, since DEC is a tool to solve all kinds of partial differential equations, this method can also be applied to solve equations in many other fields, such as Navier-Stokes equations in fluid dynamics \cite{mohamed2016}, Boltzmann equation in statistical mechanics, and Schr\"{o}dinger equation in quantum mechanics.

\section*{References}

\bibliography{mybibfile}

\end{document}